\documentstyle[prb,aps]{revtex}
\begin{document}
\draft
\twocolumn[\hsize\textwidth\columnwidth\hsize\csname @twocolumnfalse\endcsname
\title{Dynamics of SiO$_2$-glasses } 
\author{C. Oligschleger} 
\address{Institut f\"ur Physikalische und Theoretische Chemie, Universit\"at
Bonn, D-53115 Bonn, and Institut f\"ur Algorithmen und Wissenschaftliches
Rechnen, GMD -- Forschungszentrum Informationstechnik, D-53754 Sankt Augustin,
Germany}
\date{\today} 
\maketitle
\begin{abstract}
Results on the dynamics of silica are presented: vibrations and relaxations.
Using molecular dynamics, glass structures are generated by rapidly quenching
melts below the glass transition. For the local minima of the structures the
vibrational density of states is determined, the structures of the eigenmodes
are analyzed, and the influence of the single components is discussed. 
Relaxations are studied in both amorphous SiO$_2$ and  silica melts. Our main
focus is on the type of relaxation, i.e., whether the main contributions are
caused by small atomic displacements, or by bond-breaking  processes, e.g.
creation and annihilation of dangling bonds. The eigenvector of the
relaxations is found to be similar to the low-frequency one. 
\end{abstract}
]
\pagebreak

\section{Introduction}

Glasses belong to the oldest manufactured materials made by man. But despite
the long traditions and abilities in producing glasses still sufficient
information about the atomic structures in amorphous and glassy states is lacking.
Due to the absence of periodicity found in crystals both experiment and
theory are challenged to work with sophisticated methods. The technological
relevance of ceramics has led to increased efforts to resolve the
structures and to gain insight into the so-called structure-property-relationship
which can be exploited for the design of new materials. Neutron- or
x-ray-scattering provides information about the global arrangement of the
atoms via the radial distribution of the pair-correlation \cite{Wright}.
Electron spin resonance and positron annihilation is used to detect dangling
bonds in the structures \cite{Maier}. From nuclear magnetic resonance studies
one can get insight into the distribution of bond angles
\cite{Neuefeind}, however models are necessary to interpret the results.
Due to the enormous increase in computer power, simulation has become a widely
used tool to model amorphous 
structures. The applied techniques encompass molecular dynamics
(MD)\cite{Woodcock,StillingerW85,Feuston,CarP88,VKRE,TATM,TseK93,Vessal,Vollm},
Monte Carlo (MC) \cite{Grossmann,Shiff} and molecular modelling
(MM)\cite{Dean,GladdenEll}, each of which has advantages
and draw backs. Molecular modelling is successfully applied to covalently
bonded materials, e.g. SiO$_2$ \cite{Gladden} and GeO$_2$ \cite{SW}. Its main
advantage is that, without applying refined interaction potentials,
high-quality structures can be modelled which show all the experimentally
known data {\it per constructionem}. The draw-back is the need of 
information about typical bond-lengths and bond-angles of structural units
in the glass. Monte Carlo methods are applied to investigate the phase space
and its properties and to study relaxations from a global point of view
\cite{BerryB95}. The Reverse Monte Carlo (RMC) technique is often used to build
structures according to experimental information, e.g. neutron-scattering data
\cite{McGreevy}. The disadvantage of MC is that moveclasses may often be
unphysical and do not give reliable insight into the {\it microscopic} dynamics
of the systems.
Molecular dynamics is widely used to construct models of amorphous state by
rapidly quenching melts and analyzing the dynamics of the models
\cite{AllenTildesley}. Due to the
fact that high-quality interaction potentials are necessary to simulate both
structure and dynamics of the materials under consideration the results of MD
strongly depend on these potentials.
In all cases (MM, MC and MD) the typical system sizes are of the order of
several thousand atoms, which correspond to a volume of several ten thousands
cubic Angstroms. These small system sizes cannot be used to simulate large
scale effects, e.g. grain boundaries (where a grain would be of the order of a
typically available model size). 

Compared to crystalline phases amorphous materials show a richer 
dynamics which plays an important role with respect to low temperature thermal
properties \cite{Phill81,Guent81}.
Apart from the long wavelength modes (sound waves) also present in crystals,
in glasses strongly overdamped modes are observed at low frequencies.
These strongly overdamped modes lead to a strong reduction of the thermal
conductivity in glasses compared to their crystalline {\it analogues}.
These modes can also be seen in the excess specific heat at low temperatures .
typically below a few K. \cite{ZellerPohl}  In the corresponding crystals one
finds an increase of the specific heat with $T^3$ according to the Debye model
\cite{Ashcroft,Schober_c5}. The low temperature behaviour of glasses is well
described by the tunneling model assuming a constant distribution of two-level
tunneling states  \cite{AHV72,Phillips72}. 

At somewhat higher temperatures the specific heat shows a bump in $c_p/T^3$.
Using Raman \cite{Winterling} and neutron scattering \cite{Buche88} one has
deduced that thermally activated jumps over energy barriers and vibrations,
which become anharmonic at the lowest frequencies, are responsible for such a
behaviour. 
The influence of these effects is well described by the soft potential model
(SPM) \cite{KKI,IKP} in which one assumes that the dynamics of soft regions in
the structures is governed by soft potentials. The SPM is an extension of the
standard tunneling model which is included in the SPM as its low temperature
limit.
The parameters describing the softness of both the two- and one-well
potentials are distributed according to some probability distributions.
Fitting these parameters
to experimental data one finds that the regions comprising the active vibrating
atoms are confined to about 20 to 100 atoms \cite{BGGS,BGGPS}. 

Recently, computer simulations of soft sphere glasses (SSG) have supported the
view of 
soft local modes \cite{LairdS91,SLaird91} with entities comprising about 20 to
50 atoms. Similar modes  
have since been observed
in simulations of covalently bonded glasses, e.g. SiO${}_2$ \cite{Jin93} and
Se \cite{Olig93}, in metallic glasses, such as Ni-Zr \cite{Hafner94}, Pd-Si
\cite{BalloneR95}, and amorphous and quasi-crystalline Al-Zn-Mg \cite{Hafner93}
and other systems, for example amorphous ice \cite{Cho94}.  In an
earlier simulation of amorphous silicon low-frequency localized
vibrations have been observed at coordination defects \cite{Biswas88}.

Further and in addition to these periodic excitations, relaxations (aperiodic
rearrangements) are observed in ultrasonic and dielectric relaxation
experiments. \cite{RaychaudhuriHunklinger84} These processes occur 
already at low temperatures ($T>5$~K) and can be either tunneling
processes or thermally activated jumps over energy barriers.
In the SPM local relaxations are strongly correlated with local soft
vibrations, assuming a smooth transition from vibrations to relaxations.
Experimentally this is supported by the similarity of the structure factors of
both types of excitations \cite{Buche88}. 
From diffusion measurements in a metallic glass effective masses of
ten atoms have been derived for this collective motion \cite{Faupel90}.
Strong correlations between soft vibrations and hopping have been observed
in molecular dynamics simulation of the SSG \cite{SOL93} and in amorphous Se
\cite{OligschlegerS}. Both reversible and irreversible relaxations were
observed which consist of a collective hopping of groups of atoms. 
Collective hopping motions have also been observed in simulations of binary
Lennard-Jones and soft--sphere mixtures above and below the glass
transition temperature \cite{Miyag88,Wahnst91} and in amorphous Ar
after introduction of vacancies \cite{Delaye}.  Heuer and Silbey
\cite{Heuer93} systematically searched for double well potentials in a
binary model glass at $T=0$. They found a distribution of two well
potentials in qualitative agreement with the soft potential model.

In this paper we investigate the structure, vibrations and relaxations in
amorphous silica. The following chapter is devoted to computational details.
The analysis of the structures is given in section \ref{structure}.
The vibrations and related quantities are shown in section \ref{spectra}.
The results of a detailed analysis of the modes similar to the one given by
Taraskin and Elliott \cite{TarEll97} are presented in section
\ref{mode_analysis}. The study of relaxations in silica glasses and melts
(section \ref{relaxations}) is the most important part of the work presented
here. To our knowledge this is the first extensive such study.
As done in previous work on SSG and Se, the relaxations are analyzed with respect
to jump-length, localization and element specific contributions to the atomic
rearrangements. In the last section the results are discussed and conclusions
are drawn.

\section{Computational details}\label{comput}

\subsection{SiO$_2$ and its potential}

$\rm{SiO_2}$ exists in many different crystalline allotropes (e.g. $\alpha$-
and $\beta$-quartz, high- and low-cristobalite, tridymite, keatite, coesite and
stishovite) \cite{Wy82}. It is known to be a strong glass former
\cite{Wright,Angell88}.

The atomic interaction potential used in our simulation was fitted
by Vashishta et al. \cite{VKRE} in order to reproduce structural and dynamical
properties of both crystalline and glassy phases.
The potential consists of two-body and three-body-terms. The two-particle
interaction contains a long-range Coulomb part which we treat by an Ewald
summation \cite{ewald}, and includes terms for steric repulsion of the
particles and polarizability of the atoms.

The three-body-interaction is only relevant for Si-O-Si and O-Si-O units, for
the other possible combinations ( Si-O-O, O-Si-Si, O-O-O, and Si-Si-Si) it is
neglected. The three-particle terms favour the tetrahedral angle at the silicon
(O-Si-O) and an angle $\Theta_0 \approx  140^\circ$ at the corner sharing
oxygen (Si-O-Si). For details see Ref. \onlinecite{VKRE}. 

\subsection{Structures and Relaxations}

To produce silica glasses we quenched equilibrated melts below the glass
transition temperature $T_g$, the number density is kept constant during all
MD-runs and corresponds to a density  $\rho = 2.20$~g/cm$^3$.
An estimate of the glass-transition temperature was obtained by
calculating the diffusion constant, D, of the system as a function of
temperature using the relation
\begin{equation}
D_{\alpha} = \lim_{t \rightarrow \infty} \frac{1}{6t}
<\mid {\bf R}_{\alpha}(0) - {\bf R}_{\alpha}(t) \mid^{2}> \; ,
\end{equation}
where ${\bf R}_{\alpha}(t)$ is the time-dependent position vector of a particle
of type $\alpha \in \lbrace$Si, O$\rbrace$ and $<...>$ denotes the
configurational average. 

In Fig.\ \ref{diff} the temperature dependent self diffusion constant for both
silicon and oxygen shows a rapid drop at a temperature $T_0 \approx 2500$~K.
This temperature $T_0$ can be taken as a upper limit for the glass
transition temperature $T_g$, since below $T_{0}$ the {\it diffusive} motion of
the particles is effectively frozen out. In the high temperature limit the
ratio between the oxygen and silicon diffusion constant is about 1.2 which is
less than one would expect from the mass ratio $m_{\rm Si}/m_{\rm O} = 1.75$
if the different species diffuse according to their respective momentum
distribution. But this effect might be
explained by the different ``self''-interaction term in the Si-Si- and
O-O-interaction.
In comparison to experiment (where one measures $T_g \approx 1500$~K
\cite{Angell88}) the calculated glass transition temperature is much too high.
This problem is common to all computer simulations and caused by too high
quench rates, which exceed the experimental ones by many orders of magnitude
\cite{Vollm}.

We used constant quench rates of $\dot T < 3\cdot 10^{13}{\rm K}/{\rm
s}$ and quenched the glasses in two steps: first we quenched the systems from
$T = 3500$~K to 1700~K, monitored relaxations and jumps in a temperature
regime well above and below $T_g$, and stored the atomic positions of 
the detected new minima. In a next step the temperature of these glasses are
quenched with the same rate to 270~K. Again we monitored the relaxations and
stored the atomic coordinates. We have generated 14 glasses with $N=1944$ atoms
each. To study system size effects we have additionally constructed 10 smaller
glasses with $N=576$. The time step in our simulations is about 1.2~fs. Surface
effects are reduced by applying periodic boundary conditions. During
equilibration we used the NVT ensemble. 

The resulting structures are used to analyse relaxations both around the
glass transition  temperature $T_g \approx 2500$~K and well below. 
To detect the relaxations in the course of the MD runs we monitored the atomic
displacements, 
\begin{equation}\label{eq:1}
\Delta R^2(t) = \sum_n({\bf R}_n(t) - {\bf R}_n^i)^2
\end{equation}
where ${\bf R}_n(t)$ is the position vector of atom $n$ and ${\bf R}_n^i$
gives its position in minimum $i$ of the potential energy surface.
If the total displacement of the atoms exceeded a cut-off value, and the
residence time of the atoms in the new positions also exceeded a minimal period
of at least three times the period of a typical soft vibrational mode, the new
positions of the particles were accepted as a new minimum configuration. The
cut-offs of displacement and resident time, respectively, are chosen to avoid
spurious minima.
Relaxations were observed by monitoring the total displacement with respect to
the initial minimum, see Eq.\ \ref{eq:1}, the corresponding minima on the
potential energy surface, which are identified by a final quench to $T= 0$~K,
were stored and analysed.

To study the relaxations in glasses at elevated temperatures, the quenched 
glasses were heated  to temperatures between 270~K and 1670~K. At each
temperature the glasses were observed for up to 250000 time-steps,
corresponding to about  0.3~ns. 
To keep the temperature constant we averaged $T(t_i)$ over a period of 20 time
steps [$T_{av} = \frac{1}{20}\sum_{i=m}^{m+20} T(t_i)$]and scaled the
velocities after each period by $\sqrt{T_0/T_{av}}$ with $T_0$ being the
``desired'' temperature and $T_{av} =$ the averaged temperature. This has the
effect that potentially relaxing atoms are not ``slowed down'' by reducing
their velocities.
                               
\subsection{Vibrations}

After a final quench of the low temperature configurations to 0~K by a steepest
descent/conjugate gradient algorithm \cite{hsl} the minima of the potential
hypersurface are identified. In these minima, we calculated the dynamic matrix,
diagonalized it and determined its eigenvalues, the latter correspond to the
squares ($\omega^2$) of the eigenfrequencies. \cite{Lapack}
The numerically exact minimization of the
potential energy prevents the occurrence of spurious unstable modes
\cite{hsl,Fletcher}.
The elements of the dynamic matrix are given by
\begin{equation}
D^{mn}_{\alpha\beta} = \frac{1}{\sqrt {M_n M_m}}
\frac{\partial^2 U(|{\bf R}^m - {\bf R}^n|)}
{\partial R^m_\alpha \partial R^n_\beta} 
\end{equation}
which are the mass-weighted second derivatives of the potential energy $U$ with
respect to the atomic positions. 
The frequency spectra are calculated from the
frequencies of the $3N-3$ vibrational modes $\sigma$ as
\begin{equation}\label{DOSeq}
Z(\nu) = \left< \frac{1}{3N-3} \sum_\sigma \delta (\nu -\nu^\sigma)
         \right>
\end{equation}
where $\delta $ is the discretized $\delta$-function
and $\left<\ldots \right>$ stands for the averaging over configurations.
Due the limited system size the spectrum is cut-off at small frequencies. To
close this gap we calculated the Debye spectrum
\begin{equation}\label{Deb_eq}
Z_{\rm{Debye}} = \frac{3}{\nu_D^3} \nu^2 
\end{equation}
with
\begin{equation}
\nu_D = \bar{c} \left( \frac{3 N}{4\pi V} \right)^{1/3}
\end{equation}
and the average sound velocity $\bar{c}$ given in terms of the longitudinal
and transverse velocities $c_\ell$ and $c_t$. These are calculated from the
elastic constants of the glass by the usual relation 
$c_\ell = \sqrt{c_{11}/\rho}$ and $c_t =\sqrt{c_{44}/\rho}$ where we 
employ the elastic isotropy of the glass.

We calculated the elastic constants from the change in potential energy, 
$\Delta E$ under an applied strain 
\begin{equation}
 R^m_\alpha \to R^m_\alpha + \sum_\beta \epsilon_{\alpha\beta} R^m_\beta .
\end{equation}
\begin{equation}
\Delta E = - \sum_{\alpha\beta} P_{\alpha\beta} \epsilon_{\alpha\beta}
           + \frac{V}{2} \sum_{\alpha\beta\gamma\delta} 
             \epsilon_{\alpha\beta}   
                 C_{\alpha\beta\gamma\delta} \epsilon_{\gamma\delta}
           + \frac{1}{2} \sum_{\alpha\beta\gamma} P_{\alpha\beta}
                  \epsilon_{\alpha\gamma}\epsilon_{\gamma\beta} .
\end{equation}
Here the first term accounts for the work done against the forces for an
ensemble which is not in equilibrium against volume changes where
$P_{\alpha\beta}$ is the virial of the forces. The third term is a correction
for the volume change
under a finite shear in such a lattice and the $C_{\alpha\beta\gamma\delta}$
are the elastic constants ($c_{11} = C_{1111}, \ c_{44} = C_{2323}$).

We found for the glass the sound velocities $c_\ell = 6100{\rm m/s}$  and 
$c_t = 4600{\rm m/s}$ which are about 10$\%$ and 20$\%$ higher than the 
experimental values. Taking the finite size of the ensemble into
account this implies that the lowest frequency sound wave which could be observed 
for the largest sample (N=1944 atoms) is about $\nu \approx 1.5$~THz.

\section{Structure of Silica-glasses}\label{structure}

In Table \ref{TableI} the distribution of the coordination numbers CN for
the single components at different temperatures is given. If the atomic
distances $r_{ij} < 2.3$~\AA{} (this distance corresponds to the minimum in the
pair-correlation function, see below) then atoms $i$ and $j$ are neighbours.
One can
clearly observe a reduction of the fraction of ``coordiantion defects'', i.e.
atoms with another coordination number than the most common one, with decreasing
temperature. During cooling the local topology of the atoms improves due to
relaxation of the structures.                
In the liquid phase at $T=3500$~K nearly 10$\%$ of the particles are under- or
over-coordinated. Whereas equal fractions of oxygen atoms are under- and
over-coordinated, nearly no over-coordinated silicon is found. 
However, in the glassy state at $T=870$~K the fractions of under-coordinated Si
and O are reduced to less than 5$\%$ and less than 3$\%$, respectively.
Over-coordinated oxygen atoms have nearly disappeared, and nearly no
over-coordinated silicon is found. In the glasses quenched from 
$T=270$~K to $T=0$~K the total fraction of defects is even less than
3$\%$. From experiments it is known that in silica coordination defects
exist in which the silicon is only surrounded by three oxygen (oxygen vacancy)
\cite{Griscom} the number of these defects is enhanced by irradiation. Our numbers
of coordination defects are comparable to other models and simulations.
In molecular dynamics simulation of silica glasses the numbers of defects can
vary between less than 2$\%$ \cite{Feuston,Vollm} and 6$\%$ to 8$\%$ as pointed
out in Ref.\ \onlinecite{Feuston} and references therein, the effective
coordination numbers strongly depend on the applied interaction potentials
\cite{Feuston} and quench-rates \cite{Vollm}. 
Using molecular modelling techniques vitreous structures without any
defects can be constructed {\it per definitionem} \cite{Gladden,SW}. Typically
these structures are cluster models, i.e. without periodic boundary conditions.
                               
Quantities well-suited to gain insight into structural correlations and 
comparing simulations and experiments are the total and partial
pair-correlation functions.  We take the usual definition of the partial
pair-distribution functions $g_{\alpha\beta}(r)$ 
\begin{equation}               
<n_{\alpha\beta}(r)> \Delta r = 4\pi r^2 \Delta r \rho_N c_{\beta}
g_{\alpha\beta}(r)             
\end{equation}                 
where $n_{\alpha\beta}(r)\Delta r$ is the number of particles of species
$\beta$ in a shell of thickness $\Delta r$ and radius $r$ around a particle 
of species $\alpha$ and $<...>$ denotes the configurational average.
$\rho_N = N/V$ is the number density and $c_{\beta} =
N_{\beta}/N$ the concentration of species $\beta$. The partial
pair-correlation functions obtained at $T=270$~K shown in Fig. \ref{gofr}(b-d)
are used to assign the peaks in the total $g(r)$, Fig. \ref{gofr}(a), to the
partial contributions. From the integral over the first maximum of the
pair-correlation functions one can deduce average 
coordination numbers: each silicon has 3.96 oxygen neighbors. The coordination
numbers of the O-O-correlation reaches a number 6 for the range
from 0.25~nm to 0.3~nm. For Si-Si we observe a value of 4 in the range from
0.28~nm to 0.35~nm.           
From these partial pair-correlations one can deduce that the peak at
0.42~nm with a shoulder at about 0.35~nm is a superposition of the
second peak in the Si-O-contribution and the shoulder of the second O-O-peak.
The fifth peak in the pair-distribution function at 0.5~nm in Fig. \ref{gofr}
appears to be a superposition of the second peaks of the O-O- and
Si-Si-pair correlations.
                               
In Fig. \ref{dofr} the pair-distribution function $d(r)=4\pi r \rho_N
(g(r)-1)$ averaged over an ensemble of 14 glasses with $N=1944$ atoms is
shown.Together with the result of an electron scattering experiment
\cite{HeiMa}. Up to For $r \approx 0.6$~nm, i.e. twice the average
Si-Si-distance, both curves agree very well. (The wiggles below the first peak
in the experimental curve are an artefact.)

The bond-angle distributions are plotted in Fig. \ref{angles}.
The O-Si-O distribution has its peak-position at
109$^{\circ}$ and a full width at half maximum (FWHM) of 10$^{\circ}$. Such a
narrow distribution of the tetrahedron-angle suggests
slightly distorted tetrahedra as basic structural unit. 
The Si-O-Si distribution shows a peak at
145$^{\circ}$ and a FWHM of 25$^{\circ}$. This broader angle distribution
refelcts the disorder in the silica glasses. In the O-O-Si
angle distribution the peak at 35$^{\circ}$ stems from atoms belonging to
the same tetrahedron. It is directly connected to the sharp peak
in the O-Si-O angle distribution. The broad distribution
between 80$^{\circ}$ and 180$^{\circ}$ stems from configurations where the two
O-atoms have different Si's as nearest neighbours. In the Si-Si-O angle distribution
the peak around 20$^{\circ}$ stems from two silicon atoms bonded to the same
oxygen. This peak is directly connected to the Si-O-Si distribution.
If an oxygen is not a bridging atom between two silicon atoms, but instead 
neighbouring the adatom, we find an angle between 60$^{\circ}$ and
160$^{\circ}$. The strong peak at 60$^{\circ}$ in the
O-O-O angle distribution is from triplets of O's bonded to the same
Si-atom. For a situation in which the oxygen atoms are bonded to
different Si the O-O-O angles are always larger.
For the Si-Si-Si angle we find a broad distribution between
80$^{\circ}$ and 180$^{\circ}$ with a maximum at about 105$^{\circ}$. This
might give a hint to patterns in which the silicons are surrounded by strongly
distorted ``Si-tetrahedra''.             
The distance between the next-nearest silicons in the triple is about 0.5~nm.
In triples where all Si are at their nearest neighbor distance
($\approx 0.3$~nm) we find Si-Si-Si angles of about  60$^{\circ}$ which
explains the small vicinal maximum in the distribution. This points to the
existence of 6-membered Si-O-rings in silica.
In all, the angle distribution clearly reveals corner-sharing tetrahedra of 
SiO$_4$.                       
                               
\section{Spectra and localization of vibrational modes}\label{spectra}

Using Eq.\ \ref{DOSeq} we show the vibrational density of states as $Z(\nu)$
in Fig. \ref{spec_deb} averaged over all configurations together with the
Debye spectrum, Eq. \ref{Deb_eq}. One sees a clear enhancement of the glassy
spectrum at low frequencies which reaches well beyond frequencies where
system size could be of importance. Reflecting some shortcomings of the
interaction potential the experimentally observed double peak at high
frequencies \cite{Carpenter} cannot be reproduced.
The Bose-peak, i.e. the maximum of $Z(\nu)/\nu^2$, lies at $\approx$ 1.6~THz
higher than experiment where one observes it at about 1~THz \cite{Buche86}.  
From the density of states we can calculate the vibrational specific heat
at constant volume $c_v$. In the harmonic approximation one has per atom

\begin{equation}
c_v = 3 k_B \int d\omega \left[ \left( \frac{\hbar\omega}{2 k_B} \right)^2
     / \sinh^2{(\hbar\omega / 2 k_B T)} \right] Z(\omega).
\end{equation}

The vibrational specific heat  of a perfect crystal at low temperatures is
$ \propto T^3$. It is, therefore, usual practice to plot $c_v / T^3$. In such
a plot a Debye spectrum gives a constant whereas in glasses an increase above
this constant is found. 

Fig. \ref{spec_heat} shows this behavior for the silica glasses.
The solid line shows the values gained from the spectrum of Fig. 1 which is
corrected for the finite size of the simulated glass by adding a Debye
contribution up to a frequency smaller than the lowest possible
phonon frequency. This correction amounts to a fraction of $7.7 \times 10^{-3}$
of all modes. The resulting values for the specific heat are shown by the full
line. The diamonds corresponds to experimental values \cite{ZellerPohl}. The
too high values of the sound velocities in our model cause a too high value of
$\nu_D$ which leads to a Debye contribution to $c_V$ which is too small by a
factor of $\approx$ 1.6. 

The phonon eigenvectors describe the structure of the vibration and can, e.g.
be used to determine the degree of the localization of the
vibration. There are two commonly used measures of localization, the
effective mass and the participation ratio \cite{Dean,LairdS91}.
The effective mass is given in terms of the eigenvector as
\begin{equation}\label{eff_mass}
m_{\rm{eff}}(\sigma) = m / |{\bf e^1(\sigma )}|^2 .
\end{equation}
Here we have assumed that the $3N$-dimensional unit vector of mode $\sigma$
is normalized and ${\bf e}^n(\sigma )$ stands for the vector formed from the
three components on atom $n$. Atom number 1 is chosen as the atom with the
largest displacement. $m_{\rm{eff}}/m$ is a measure for the number of
atoms which effectively carry the kinetic energy of the vibrational mode.
This definition is limited to small system sizes when the long range
tails of the modes are not too important. 
In the following we will mainly use the participation ratio:
\begin{equation}\label{part_mode}
p(\sigma ) = \left( N \sum_{n=1}^N |{\bf e}^n(\sigma )|^4 \right)^{-1} .
\end{equation} 
For a translation one has $p=1$ and for
a vibration of a single atom with all others at rest $p=1/N$. This
scaling with $1/N$ should hold for all localized modes.
Fig. \ref{part_spec} shows the participation ratios for the two system sizes
with $N = 576$ and $N=1944$ atoms.
The high-frequency modes are highly localized and their participation ratios
show the expected scaling with system size.
At the low-frequency end we observe (quasi-)localized modes. The
localization  corresponds to clusters comprising about 10 to 50 atoms. The
strongest localized modes have participation ratios of 0.012 corresponding to
an entity comprising about 8 atoms. These findings are in agreement with ones
by Jin {\it et al.} \cite{Jin93} and Taraskin and Elliott \cite{TarEll97}.  
However, their  participation ratios do not show the expected $1/N$ scaling.
This will be due to interaction effects between the modes in the larger
structures which results in an increase of the participation ratio. In the
larger systems there are more of similar frequency, and in particular more low
frequency extended phonons which will mix with the
quasi-localized modes and raise their participation ratio. Additionally this
causes an increased interaction between quasi localized modes
of similar energy, since due to de-localization these modes lose their
``splendid isolation'' and ``feel'' also other localized vibrations, too.
Due to this interaction of quasi localized modes their participation ratio is
increased. For a finite concentration $c$ of
interacting modes the scaling factor $1/N$ in Eq.\ \ref{part_mode} will be
replaced by $c$.
Such effects have also been found in a model glass
\cite{SchoberO96} where the modes are deconvoluted such as to reconstruct the
``naked'' contributions: phonons and localized vibrations.

\section{Analysis of vibrational modes}\label{mode_analysis}

To gain more insight into the dynamics of the vibrational modes we analysed
the motion of the single components and their contribution to the eigenvectors.
The averaged contribution of the oxygen to the mass-weighted eigenmodes in
comparison with the silicon is depicted in Fig. \ref{do_dsi}. This 
contribution is  given as the mean square displacement of each component
$\alpha \in \lbrace$Si,O$\rbrace$ to the eigenvectors
\begin{equation}
Z^{\alpha}(\nu) = \left< \frac{1}{3N-3} \sum_\sigma \delta (\nu -\nu^\sigma)
 \sum_{i=1}^{N_{\alpha}}(e_i^{\sigma})^2 \right>, 
\end{equation}
with $N_{\alpha}$ the number of the Si- and O-atoms, respectively.
Just below the high-frequency peak the silicon atoms contribute stronger to the
eigenvectors than the oxygen atoms, whereas the eigenmodes of the low-frequency
part of the spectrum and the high-frequency modes are clearly dominated by the
oxygen motion. 
To learn about typical motions we calculated the angle between the
displacement of atom $i$ in the modes (${\bf e}_i$), and the bonds of this 
atom to its nearest neighbors $j$, i.e. ${\bf r}_{\rm bonds}={\bf r}_{ij}$
\begin{equation}
\alpha_{e_i \rm bonds} = \arccos\left(\frac{ {\bf e}_i {\bf
r}_{\rm{bonds}}}{|{\bf e}_i||{\bf r}_{\rm bonds}|}\right) 
\end{equation}

In Fig. \ref{mode_all_angle} we show the weighted distribution of the angles
for the low-frequency modes of each glass with respect to the components Si and
O. We weighted with $e_i^2$ in order to suppress marginal contributions of
only slightly moving atoms. The contributions of both elements
are peaked at 90$^{\circ}$ which means the motion of both Si and O is
mainly perpendicular to the bonds. In the case of oxygen this peak is very
sharp, whereas the silicon atoms have stronger contributions to either smaller
or larger angles, this indicates that the silicon atoms have 
contributions parallel to their bonds.
Following the work by Taraskin and Elliott \cite{TarEll97} we calculated the
projections of the vibrations of the structural subunits, such as the
SiO$_4$-tetrahedra and the SiOSi-units, onto the symmetry vibrational
coordinates, i.e. internal vibrations, such as symmetric and asymmetric
stretching, bending modes, and rigid body motions, namely rotations and
translations. The translations of the sub-units are neglible ($<1\%$) in
comparison to rotations and internal vibrations. To calculate the internal
motions the center of these subunits
are fixed, i.e. the vibrations of the surrounding atoms are shifted relative to
the central atoms. The remaining motions can be either rigid rotations or
internal vibrations, such as stretching or bending modes. In Fig.
\ref{SiOSi_analysis} we show the contributions of the projections onto the
SiOSi subunits to the total DOS. The modes in the low-frequency range have a
strong rotational character, 
whereas in the high-frequency peak an asymmetric stretching mode describes
well the typical vibrations. In the mid part of the spectrum symmetric
bending and with less extension also symmetric stretching contribute to the
modes. Projecting the modes onto the symmetry vibrations of the SiO$_4$
structural subunits (see Fig. \ref{SiO4_analysis}) we find the high-frequency
peak to be described well by both asymmetric (strong contribution) and
symmetric (weak contribution) stretching modes. 
Since in the experimental DOS one observes a double peak \cite{Carpenter}, we can speculate that
our interaction potential gives a too low concentration of symmetric stretching
modes and a too high concentration of asymmetric modes. The center of the
spectrum has
contributions from bending modes with F$_2$ and E symmetry. The low-frequency
side of the spectrum is again dominated by rotations. This is in agreement with
the observation that the atoms move perpendicularly to their bonds. 

\section{Relaxations above and below the glass transition temperature}
\label{relaxations}

At the start of the simulations the temperature of the configurations is
about 3500~K. Cooling with a quench rate of about $\dot T < 3\cdot10^{13}~\frac
{\rm K}{\rm s}$ the structures cooled to $T = 1700$~K after 50000~time steps
$\approx 60$~ps. During cooling we observe initially a rapid succession of
strong relaxations.  After about $20$~ps and at a temperature of circa 2800~K (
slightly above T$_g$), the activity of the structures slows down,
i.e. jumps become less frequent and the systems become temporarily trapped in
metastable regions of configuration space.

During cooling several new local minima of the potential energy surface are
visited by the ``MD-walker'' via jumps over energy barriers. To determine the
localization of the hopping processes and to test the SPM assumptions, we
calculated similar to Eq. \ref{eff_mass} the effective mass of the relaxation:
\begin{equation} \label{eq:2}
M_{\rm eff} = m_{\rm max}\cdot\frac{(\Delta R)^2}{|\Delta {\bf R}^2_{\rm max}|}
\end {equation}
where $(\Delta R)^2=\sum_n({\bf R}_n^i-{\bf R}_n^f)^2$
is the square of the total distance $\Delta R$ between two successive minimum
configurations (called ``initial'' and ``final'' positions of the jump).
${\bf R}_n^{i,f}$ denotes the respective
initial and final positions of atom $n$,
$|\Delta {\bf R}^2_{\rm max}|$ is the maximal distance a
single atom jumps in this relaxation, and $m_{\rm max}$ is the mass of the
farthest jumping atom. Similarly to Eq. \ref{part_mode} we calculated the
participation ratio $p_{\Delta R}$
to determine the
localization of the hopping processes,
\begin{equation} \label{eq:3}
p_{\Delta R} = \frac{\Delta R^4}{N\sum_n({\bf R}_n^i-{\bf R}_n^f)^4}
\end {equation}

where ${\bf R}_n^i$ and ${\bf R}_n^f$ denote the initial and final position
of atom $n$ in the relaxation and $\Delta R$ is the total jump-length. The
participation ratio has the value $n/N$ if $n$ atoms are involved in the
hopping process, and $p_{\Delta R} = 1$ if all atoms contribute equally to the
relaxation. We find that the participation ratio of jumps grows
roughly linearly with the jump distance, Fig.\ \ref{SiO2_pr}. Comparing the
results for the two temperature intervals one clearly observes an increase of
both $\Delta R$ and $p_{\Delta R}$ with temperature.
The gap between the two temperature regimes is an artifact of the cut-off
procedure used for the higher temperatures.

To study relaxations in a ``metastable equilibrium'' we heated the quenched glasses
to $T = 270$, 870 and 1670~K.
For the jumps of the glasses which are heated from $T = 0$~K to the desired
temperature  we monitored again the minima visited in course of the
MD runs and determined the corresponding jumps and their participation ratios.  

As example we plotted the time evolution of the energy and the displacement in
Fig. \ref{T_hist} for the glasses at $T=870$~K. In the low-temperature regime
we can follow an overall relaxation of the structures towards more stable
configurations. This trend becomes even stronger in the higher temperature
regime. 
The ensemble-averaged participation ratios for the equilibrated
temperatures (equilibrated during heating) show a slight temperature
dependence, $p_{\Delta R} = 0.020 \pm 0.008, 0.057 \pm 0.035, 0.061 \pm 0.036$
for $T =$ 270, 870, and 1670~K, respectively. Quite in contrast though, the
jump-lengths increase linearly with temperature, and we find $\Delta R
=1.81\pm0.75$, $4.72\pm1.38$, and $8.27\pm2.49$\AA~{} for $T = 270$, 870, and
1670~K, respectively. The typical ``resident time'' per minimum is about 15~ps.
In studies of amorphous Se the resident times per minimum is found to be about
30~ps at 0.05 $T_g$. \cite{Olig_unpub}     

To identify active regions we calculated a correlation between the observed
relaxations
\begin{equation}\label{eq:4}
c_{RR^{\prime}} = \frac{\sum_n\Delta R_n\Delta R^{\prime}_n}
{\Delta R \Delta R^{\prime}}.
\end{equation}

Since we calculated the product of the absolute values $\Delta R$ and not the
scalar product between the two jump-vectors  ${\bf \Delta R}$ and
${\bf \Delta R}^{\prime}$, we also observed {\it active} regions where almost
the same atoms contributing to a relaxation ${\bf \Delta R}^{\prime}$
jump {\it perpendicularly} to the previous relaxation $\Delta R$. In that case
$c_{RR^{\prime}}$ is close to 1. If the jumps are uncorrelated
$c_{RR^{\prime}}$ becomes $n/N$, with $n$ the typical number of
atoms contributing to the relaxations, and $N$ the total number of atoms in the
configuration.

Calculation of the correlation between successive jumps (see Eq.\ \ref{eq:4})
at the lowest considered temperature ($T=270$~K) yields values of
0.585$\pm$0.314. Astonishingly, for the glasses at $T=870$ and 1670~K very
similar values (0.654$\pm$0.089 and 0.666$\pm$0.082, respectively) are
observed. The broader distribution of values in the low-temperature simulation
may be explained by the observation that at the lowest temperature successive
relaxations are either reversible jumps (high correlations) or jumps where
different parts of the active regions of the systems contribute to the
relaxations at different times (low correlations). At higher $T$ these local
regions aggregate to a larger complex where the single entities are not
resolved in the jump-process. This can explain the relatively high correlations
between the jumps at higher $T$ where several ``active'' regions in the glasses
jump at the same time. However, at the higher temperatures successive jumps are
not reversible, but the same (local) regions are activated throughout the
observation time. 

To see whether there exists a typical ``relaxation-mode'' we calculated the
projection of the relaxations observed at $T=870$~K and 1670~K onto the bonds
of the atoms, Fig. \ref{Proj_relax_bond}.

\begin{equation}
\alpha_{r_i \rm bonds} = \arccos\left(\frac{ {\bf\Delta R}_i {\bf r}_{\rm
{bonds}}}
{|{\bf \Delta R}_i||{\bf r}_{\rm bonds}|}\right)
\end{equation}

where ${\bf \Delta R}_i$ is the relaxation vector of atom $i$.   
The distribution is Gaussian in the case of the oxygen, and centered around
90$^{\circ}$ indicating that the O-atoms relax mainly perpendicularly to the
bonds. For the silicon atoms we observe a peak at low angles which reflects the
fact that the Si-atoms relax mainly in direction of one bond. 
Since a silicon is surrounded tetrahedrally by the O-atoms, the projection onto
the remaining bonds will give the tetrahedron angle at about 109$^{\circ}$,
which is in satisfying agreement with the peak at about 105$^{\circ}$. Note
that the ratio between the peaks at 5$^{\circ}$ and 105$^{\circ}$ is about 1:3
which is the ratio one would expect in a tetrahedron if the center atom moves
parallel to {\it one} of its bonds. In analogy to the mode-analysis we
calculated the  projections of the relaxations onto the symmetry modes of
SiOSi- and SiO$_4$-subunits. Since in this study it has been important to
consider the ``real'' contributions of the atoms and not only the ``relative''
displacements as in the case of the eigenmodes we do not subtract the
displacement vector of the central atom. In the case of the SiO$_4$-subunits
and relaxations at $T = 870$~K and 1670~K we do find nearly no translations of
the subunit worth mentioning and negligible
contributions for symmetric stretchings of the oxygen atoms; 0.21 of the
relaxations of the oxygen atoms stem from asymmetric stretching, 0.40 of the
jumps are due to bendings in F$_2$-symmetry, 0.07 are bending with E-symmetry
and 0.32 of the oxygen motions come from rotations.     
Considering the SiOSi-subunits we do not record a considerable amount of 
translations of the SiOSi-pattern; and only 0.037 of the Si-relaxations are
due to symmetric stretching. The contribution of asymmetric stretching is about
0.26, symmetric bendings contribute at about 0.23 to the jumps, and 0.47 come
from rotations. The strong contributions from bending and rotations for Si and
O are in agreement with the picture that the typical relaxation is
perpendicular to the bonds. For the relaxations of silicons (in the
SiOSi-subunits) we find that about 30$\%$ of the relaxations can be ascribed to
stretching (3.7$\%$ symmetric and 26$\%$ asymmetric), for the oxygen
relaxations in the SiO$_4$-units only 21$\%$ of the relaxations can be
described by asymmetric stretching. 

The investigation of relaxations with respect to bond-changes (creation and
annihilation), where a bond is cut off at 2.3~\AA, at constant temperatures
shows that at $T = 270$~K the main contribution to a relaxation is due to small
changes of the positions of the atoms, bond-breaking occurs only in a few
cases. At $T=$ 870~K about $0.1\%$ of the bond situation is altered by
relaxations from one minimum to another. At the highest temperature the number
of bonds change up to $0.3\%$ per jump.

During the cooling runs we stored configurations for some of the glasses at
$T=870$~K and $T=270$~K. After an equilibration time of 60~ps we started
relaxation runs for these ``young glasses'' at $T= 870$~K and $T=270$~K,
respectively. Again relaxations were monitored for about 0.3~ns.
       
The participation ratios averaged over the ensembles for the equilibrated
temperatures $T=270$~K (4 glasses) and $T=870$~K (2 glasses) are $p_{\Delta R}
= 0.020 \pm 0.011$ and $p_{\Delta R} = 0.047 \pm 0.030$, respectively. 
The jump-lengths increase with temperature, we find $\Delta R =2.05\pm0.77$\AA,
$5.86\pm2.15$\AA~{}  again for $T = 270$~K and 870~K, respectively. Now the
typical ``resident time'' per minimum is about 10~ps.
Note that the jump-lengths of the ``young'' glasses are larger and the
resident time is smaller than in the corresponding ``older'' glasses.  
These results although based on weaker statistics may point to the fact that
the ``young glasses'' are more active in terms of longer jump-distances and
shorter relaxation-times than those glasses aged for a longer time.

\section{Discussion and conclusion}  

We have performed molecular dynamics calculations on silica glasses and melts.
The amorphous structures were generated by rapid quenches of melts.
The analysis of the glasses show a satisfying agreement with experimental
results for the pair-correlation functions. The angle-distributions
reveal the existence of a network of corner-sharing tetrahedra.
During cooling we observed 
an appreciable optimization of the local topology of the structures. At
$T=3500$~K the number of dangling bonds of Si and O is about 9.4$\%$ and
4.5$\%$, respectively. Caused by structural relaxations the number of these
defects decreases to 3.9$\%$ and 2.2$\%$, respectively, at $T=0$~K. This
improvement can also be seen in the shift of the peak in the Si-Si
pair-correlation function from less than 0.3~nm in the liquid, in which the
Si-Si-distance appears as a shoulder of the O-O-peak, to 0.31~nm (at
$T=270$~K). The Si-Si-distance becomes visible as a separate peak in the total
pair-correlation function.

One main focus of our study concerns the dynamics of the structures with
respect to both
periodic motions (vibrations) and aperiodic ones (relaxations). Calculation
of the element specific contributions shows the silicon motion to be dominant
just below the high frequency modes (about 28~THz). As to the rest of the
spectra, the oxygen vibrations become the main contribution to the mode.
The participation ratios show that all modes in the high-frequency
regime are localized, comprising entities with less than 15 atoms.
At the lowest frequencies the modes ---about 0.2$\%$ of
the total spectrum--- are typically quasi-localized vibrations with
participation ratios of less than 0.22 for $N = 1944$ atoms. The effective 
numbers of atoms contributing to the quasi-localized modes range from 10 to 50
particles. The atoms strongly participating in those low-frequency modes move
preferentially perpendicularly to their bonds. Investigating the subunits of
the structures such as SiO$_4$-tetrahedra and ``SiOSi-fragments'' we find that
the motion of the low-frequency modes can be described by rotations (most
important feature) with an addition of bending type modes. This picture is in
agreement with the one of rotations of coupled tetrahedra \cite{Buche88}. The
high-frequency vibrations can be described well by stretching modes. 

We studied relaxations of SiO$_2$ structures in the melt (during cooling) and
in the glass at several temperatures both during cooling and ``equilibration''.
During cooling from 3500 to 1700~K we observed rearrangements of the structures
with participation ratios of less than 0.3 (for $N=1944$ atoms) connected to
effective masses ranging from 20 to 150 which are somewhere between clear local
relaxations and
global changes. This observation indicates that several regions of
the structure are active at the same time. Lowering the temperature (from
1700~K to 300~K) the picture of locally changing structures becomes clearer and
the participation ratio of the jumps is less than 0.1. Analysis of the
relaxations observed during the ``equilibration'' runs shows that the jumps at
low temperature ($T=270$~K) typically lead to small changes of the atomic
positions with nearly no bond-breaking processes. At higher $T$ we observe
bond-changes of the atoms (both bond-breaking and bond-creations):
At $T=870$~K the numbers of bonds changed by about 1 bond out of 1000, and at
$T = 1670$~K of up to 3 bonds out of 1000 changes per relaxation. 
These relaxations are local processes comprising typically 10-50 atoms. For the
lowest temperature we find about 10 atoms contributing strongly to the
relaxations whereas at higher $T$ (870 and 1670~K) about 20-50 atoms are
contributing. However, the jump-length increases clearly linearly with
temperature.  In the course of the MD simulations the structures relax to
energetically lower lying minima. During aging of the structures this trend
seems to slow down. Further corroboration of this lies in the observation of
longer residence times of the configurations in each minimum and smaller
jump-lengths towards the end of our simulations. 

For all temperatures below the glass-transition we observed
correlated relaxations, i.e. several active regions coupled via a few atoms
contributing to the jumps. At the lowest temperature several active regions can
be distinguished whereas these regions are ``coupled'' at higher $T$, i.e. the
atoms comprising these regions are active at the same time
\cite{SOL93,OligschlegerS}.

Analyzing motion patterns of the relaxations has shown that the oxygen atoms
contribute dominantly to the relaxations by typically small displacements
mainly perpendicular to the bonds. Nearly half of the oxygen motion is due to
bond bendings and 30$\%$ to rotations (0.32). 
The silicon atoms exhibit also jumps more closely in the direction of one of
its oxygen neighbours. This shows a more stretching-like motion of the
Si-relaxation. However, the total contribution of this motion pattern amounts
about only 0.3. The rest of the silicon motion can be attributed to rotations
(0.47) and symmetric bendings (0.23).
Comparing the motion patterns of the relaxations to the eigenmodes (vibrations)
the obvious similarity between the motion pattern found in both
low-frequency vibrations and relaxations (main contributions are from rotations
and bending) indicates a strong connection between low-frequency modes and
local relaxations and supports the assumption of the soft potential model. Such
a correlation between local relaxations and eigenmodes has also been found in
amorphous selenium. \cite{OligschlegerS}
However, the relaxations have stronger contributions from bending- and
stretching-motions than the low-frequency modes which lead us to the conclusion
that also higher-frequency modes are necessary to describe relaxations and are
acting as impetus for localized jump-processes. 

\acknowledgements
The author is grateful for many stimulating discussions with U. Buchenau and
H.R. Schober. Thanks are due to M. Gastreich, A. Hannemann, H. Putz and J.C.
Sch\"on for critical reading of the manuscript.
Funding by the Deutsche Forschungsgemeinschaft through the
SFB408 is gratefully acknowledged.

\newpage

\begin{table}
\hspace*{2.5cm} T = 3500 K \hspace*{2.6cm} T = 870 K \hspace*{3.6cm} T = 0 K
\begin{center}

\begin{tabular}{c|l|l|l|l|l|l}
\hline
 CN     & Si    &    O  &  Si    &    O  &  Si   &  O    \\
\hline
 0      &  ---  & 0.000 & ---    &   --- &  ---  & ---   \\
\hline
 1      &  ---  & 0.045 & ---    & 0.025 &  ---  & 0.022 \\
\hline
 2      & 0.000 & 0.901 & ---    & 0.972 &  ---  & 0.976 \\
\hline
 3      & 0.094 & 0.047 &  0.045 & 0.003 & 0.039 & 0.002 \\
\hline
 4      & 0.905 & 0.005 &  0.955 &   --- & 0.961 &  ---  \\
\hline
 5      & 0.001 & 0.000 & ---    &   --- &  ---  &  ---  \\
\hline
\end{tabular}
\end{center}
\caption{
Distributions of the coordination numbers ($CN$) for both silicon and oxygen at
$T=3500$~K, 870~K and 0~K, above and well below the glass transition
temperature, respectively.} 
\label{TableI}
\end{table}

\newpage

\begin{figure}

\caption{Element specific self diffusion constants in SiO$_2$ for Si
($\diamond$) and O ($+$) plotted versus temperature.}
\label{diff}
\end{figure}

\begin{figure}
\caption{Total pair correlation function g(r) (a) and the partial
pair correlation functions (b-d) in amorphous SiO$_2$ at $T = 270$~K.}
\label{gofr}
\end{figure}
\begin{figure}
\caption{Comparison between the calculated pair distribution function d(r) and 
the experimental result by D.~Heinemann using electron scattering (see Ref.
\cite{HeiMa}).}
\label{dofr}
\end{figure}


\begin{figure}
\caption{Bond angle distribution in silica glasses.
The O-Si-O distribution has the peak-position at
109$^{\circ}$ and a FWHM of 10$^{\circ}$.
The Si-O-Si distribution has the peak-position at
145$^{\circ}$ and a FWHM of 25$^{\circ}$.}
\label{angles}
\end{figure}
%
%
\begin{figure}
\caption{Density of states ${\rm Z(\nu)}$ vs. ${\rm \nu}$ obtained by 
calculation of the Fourier transform of the displacement autocorrelation
function using the equation of motion (EOM) method (solid line) and
diagonalization (dashed line). The dotted line displays the Debye DOS.}
\label{spec_deb}
\end{figure}

\begin{figure}
\caption{Specific heat ${\rm c_V}$ as ${\rm c_V/T^3}$ in units of $[\mu J/g
K^4]$  vs. $T$ in K. Plot is double-logarithmic. Solid line is the specific
heat with Debye-correction and dotted line without correction. Diamonds are
experimental values of specific heat c$_p$ by Zeller and Pohl.}
\label{spec_heat}
\end{figure}

\begin{figure}

\caption{Participation ratio of the vibrational eigenmodes for glasses with
$N=1944$ ($\diamond$) and $N=576$ ($+$) atoms.}
\label{part_spec}
\end{figure}

\begin{figure}
\caption{Averaged element specific contributions to the vibrational spectrum
for 14 glasses  with $N=1944$ atoms. Total DOS (solid line), O-contribution
(dashed line) and Si-contribution (broken line) are plotted versus frequency in
THz.} 
\label{do_dsi}
\end{figure}
%
%
\begin{figure}
\caption{Weighted averaged element specific distribution of the angles
$\alpha_{e {\rm bonds}}$ between the atomic eigenvector and the bonds for all
low-frequency modes. The full line show the angle-distribution found for
silicon, the broken line for oxygen.} 
\label{mode_all_angle}
\end{figure}

\begin{figure}
\caption{The vibrational DOS (solid line) and the partial contributions for
the projections of the relative atomic displacement onto the symmetry
coordinates of the SiOSi-subunits with the contributions from rotations
(dashed-dotted  line), symmetric stretching (broken line), asymmetric
stretching (dashed line) and from symmetric bending (dotted line) plotted
versus frequency in THz.}
\label{SiOSi_analysis}
\end{figure}

\begin{figure}
\caption{The DOS (solid line) and the partial contributions for the projections
of the relative atomic displacements onto the symmetry coordinates of the
SiO$_4$-subunits with the contributions from rotations (dashed-dotted line),
symmetric stretching (broken line), asymmetric stretching (dashed line) and
from bending in E-symmetry (dotted line) and in F$_2$-symmetry
(dashed-dot-dotted line) plotted versus frequency in THz.} 
\label{SiO4_analysis}
\end{figure}

\begin{figure}
\caption{Displacement of O vs. displacement of Si in single jumps.
The line shows the calculated O displacement if one scales $\Delta R^{Si}$
by the square root of the mass ratio O to Si and by the ratio of the atomic
numbers.} 
\label{ro_rsi}
\end{figure}
%
%
\begin{figure}

\caption{Participation ratio for the quenching regimes vs. displacement. The
line is a guide to the eye.}
\label{SiO2_pr}
\end{figure}
\begin{figure}
%
\caption{Change of potential energies per atom averaged over an 
ensemble of 8 MD-runs at $T=870$~K (a). Total displacements
${\rm \Delta R^2}$ over an ensemble of 8 MD-runs at $T=870$~K (b).}
\label{T_hist}
\end{figure}

\begin{figure}

\caption{Distribution of the angle $\alpha_{{\rm r bonds}}$ between the
relaxations $r$ and the bonds of the atoms, full line Si-relaxation,
dashed line O-relaxation.}
\label{Proj_relax_bond}
\end{figure}
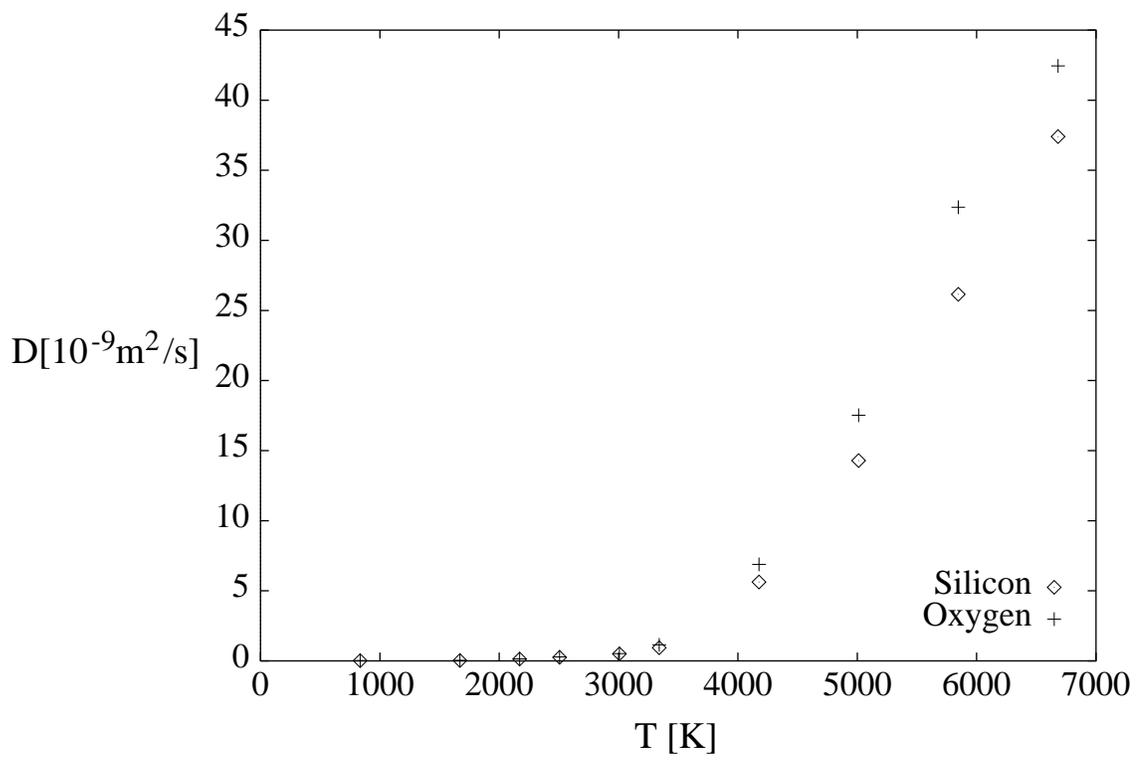
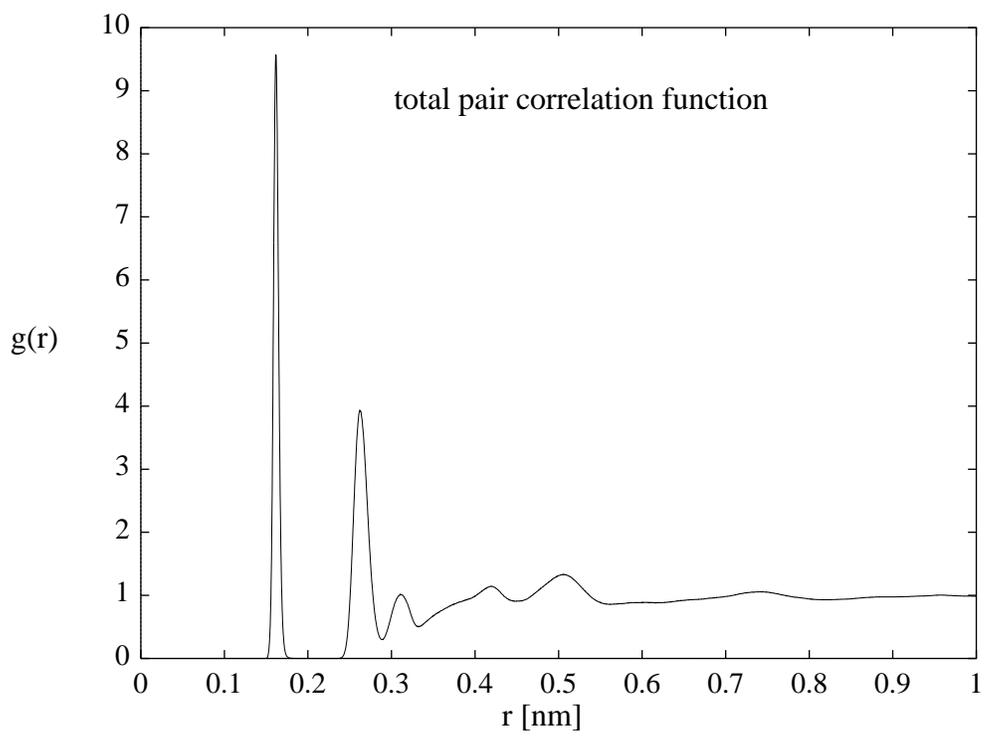
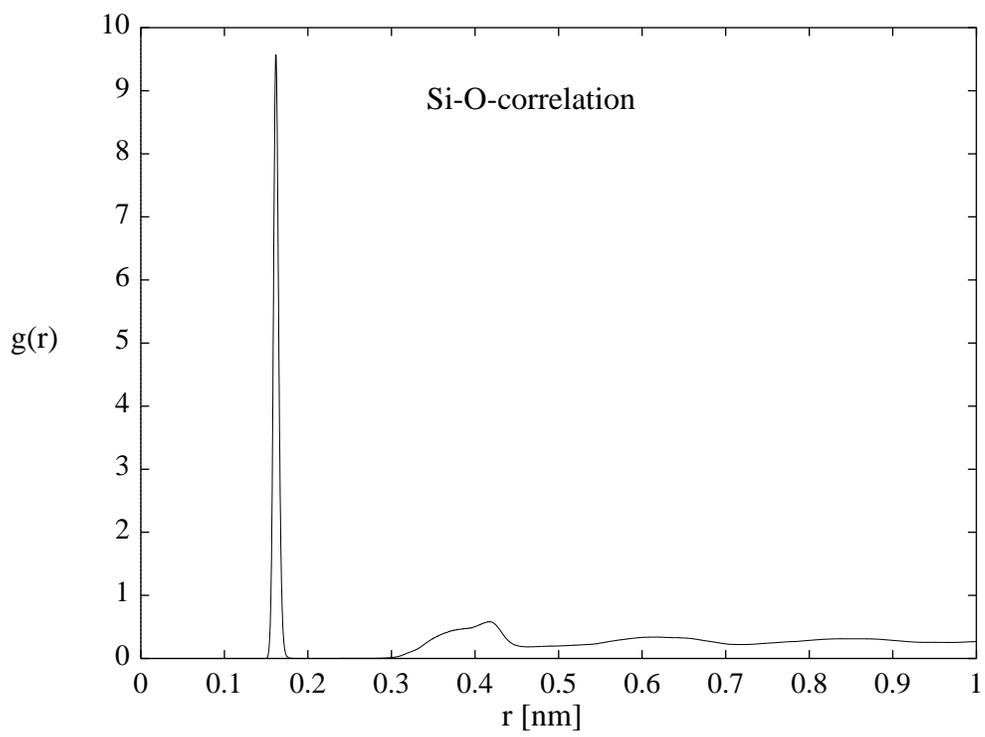
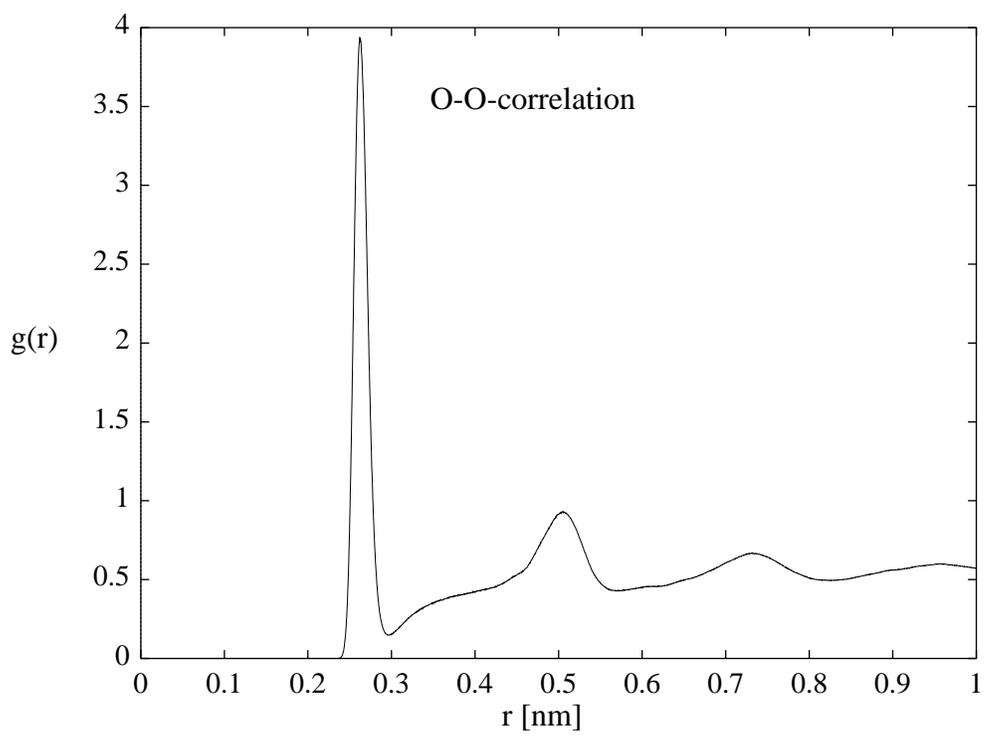
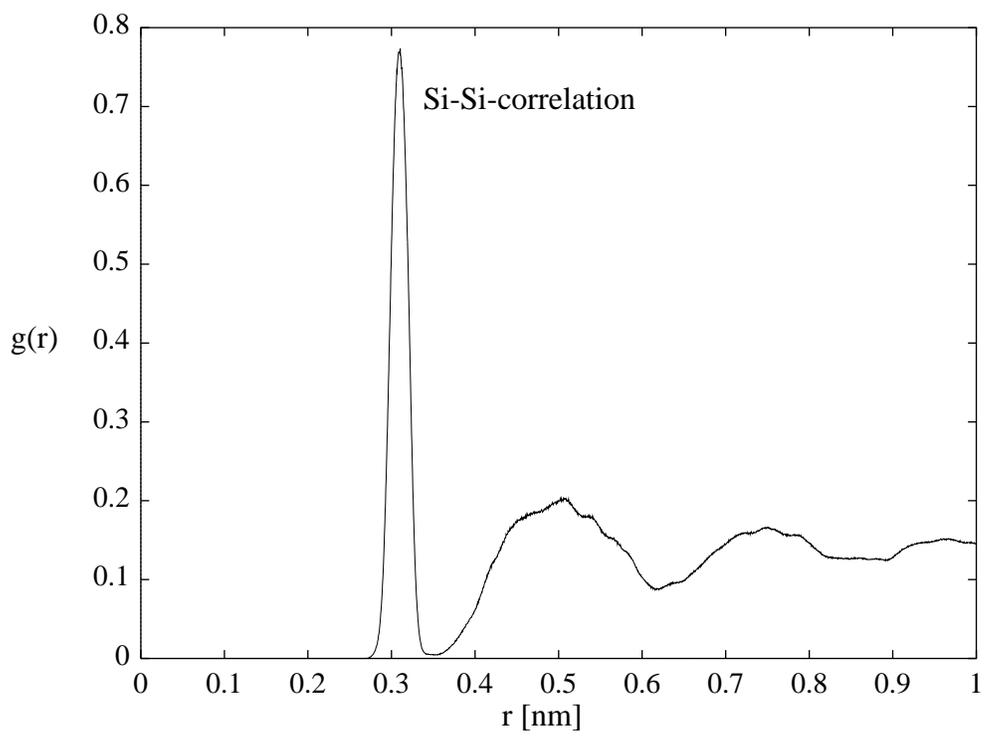
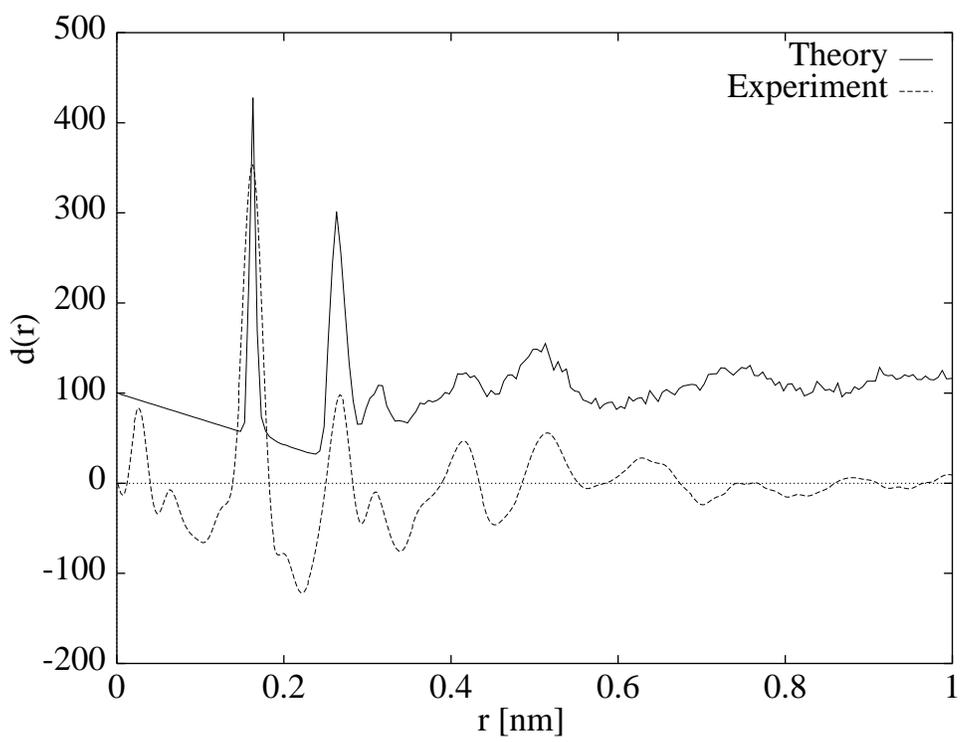
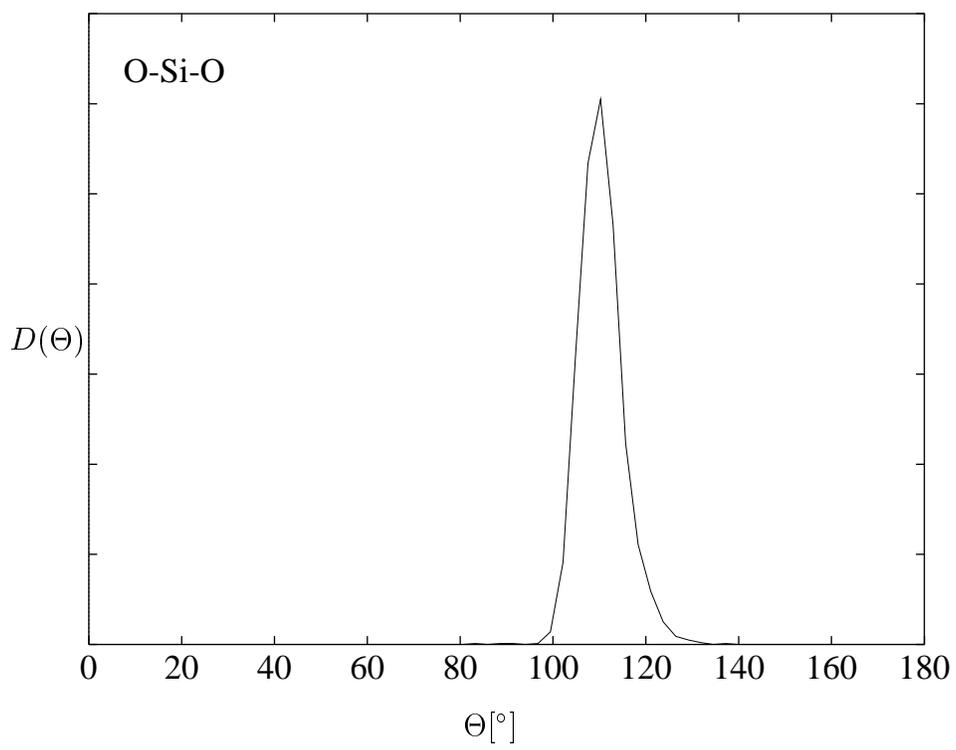
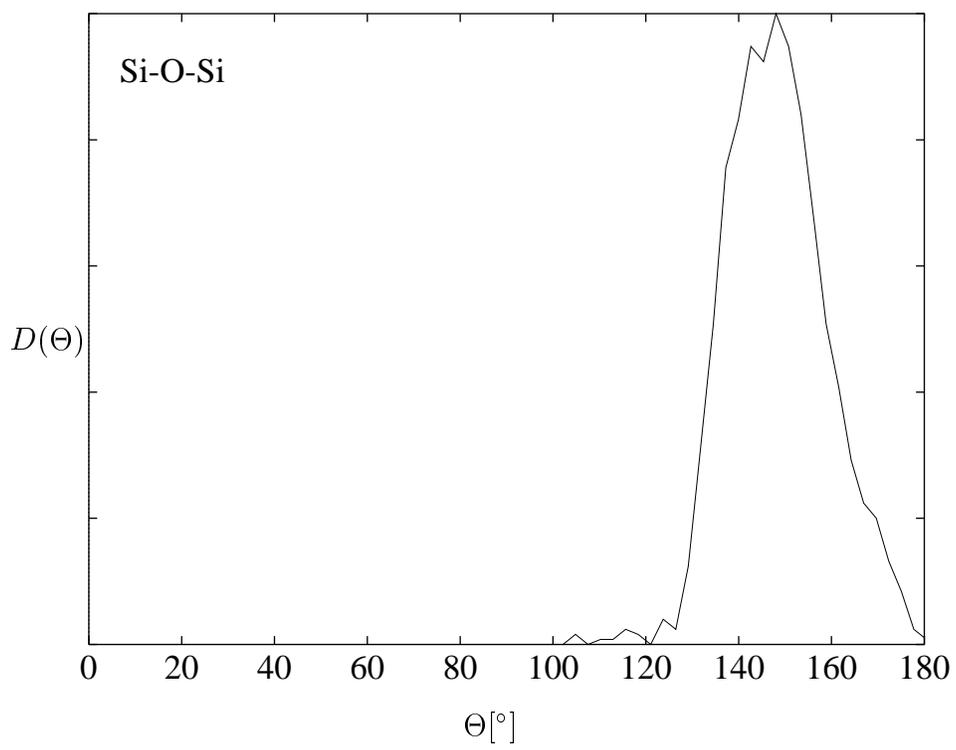
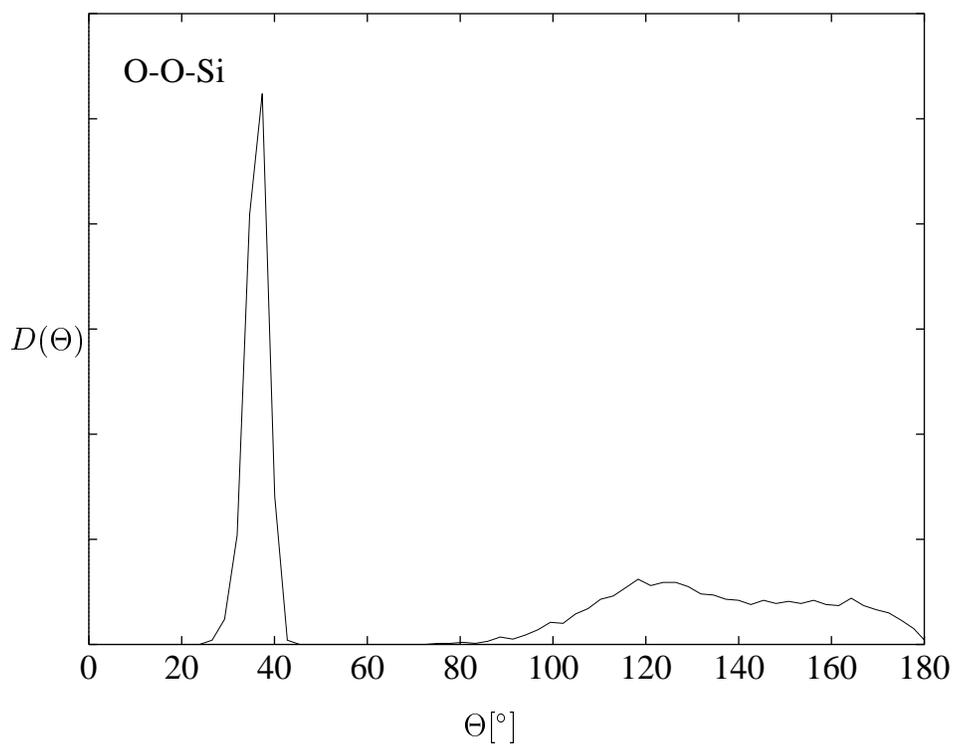
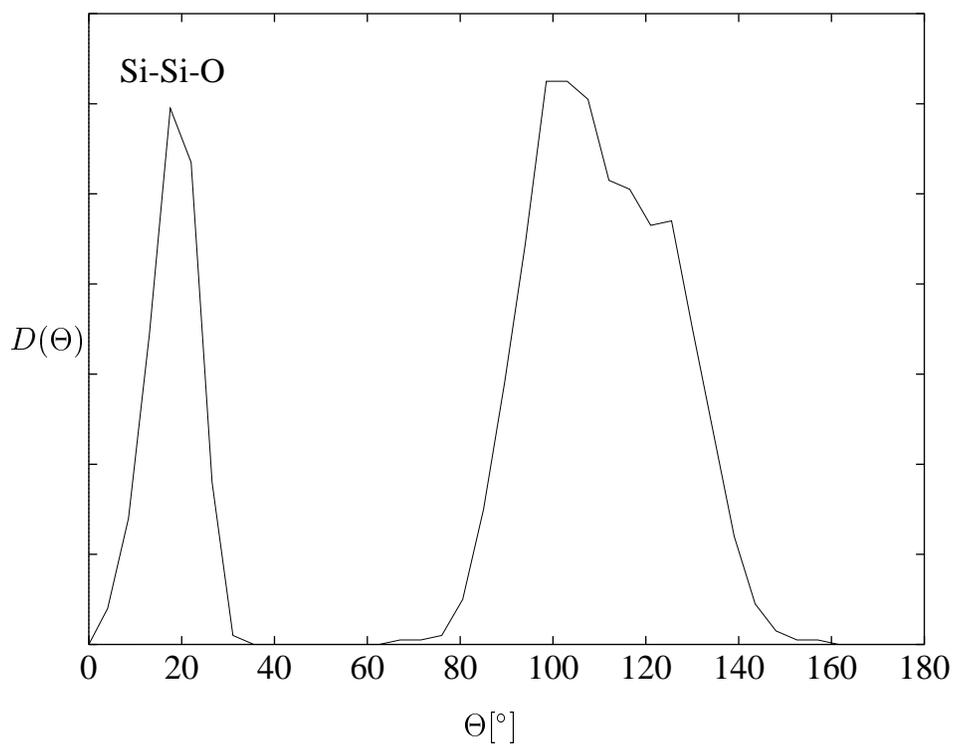
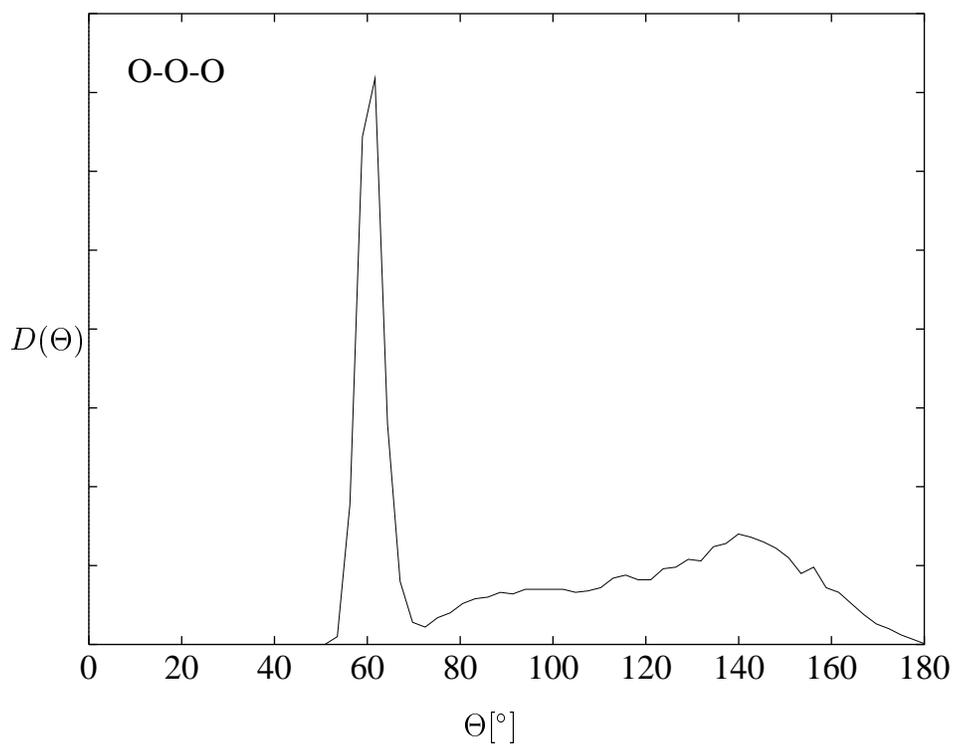
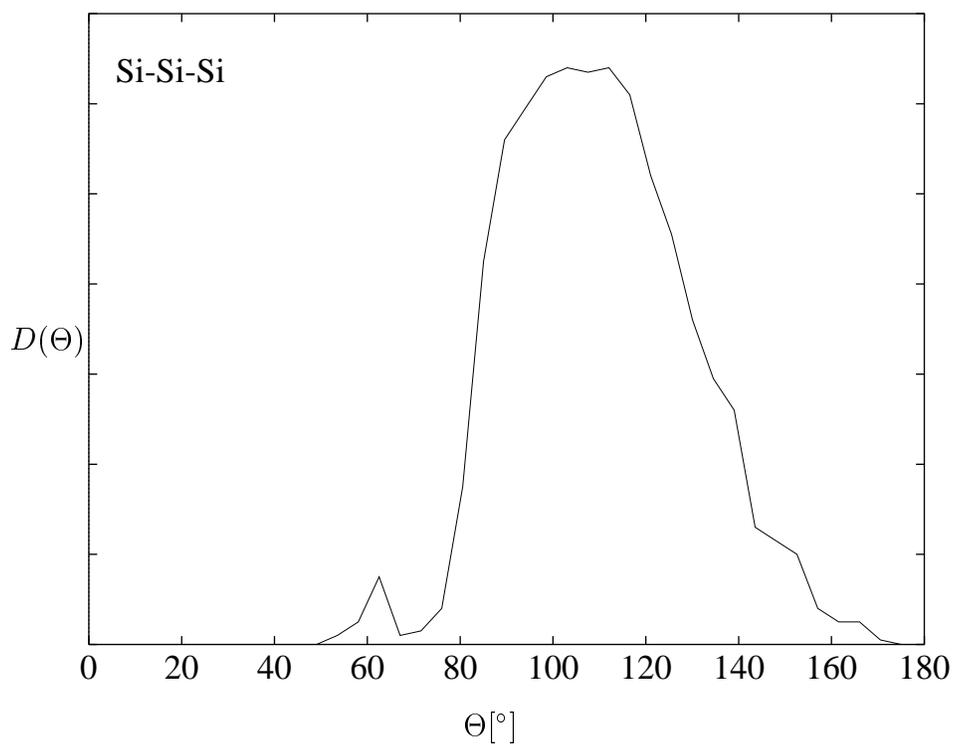
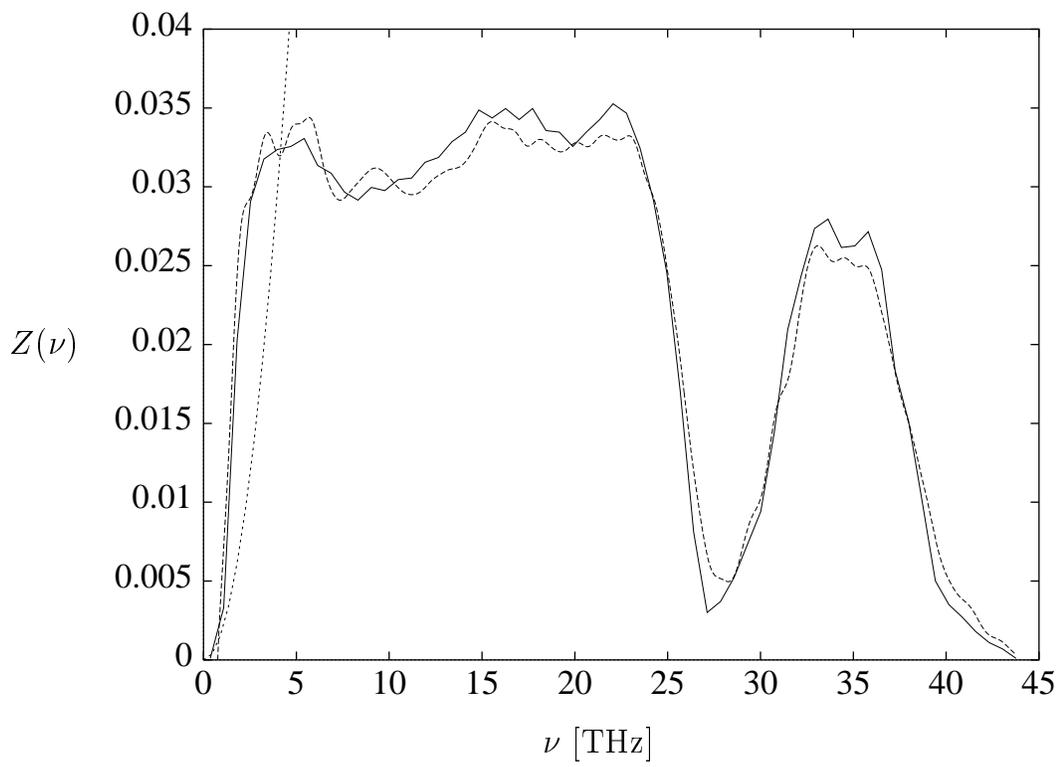
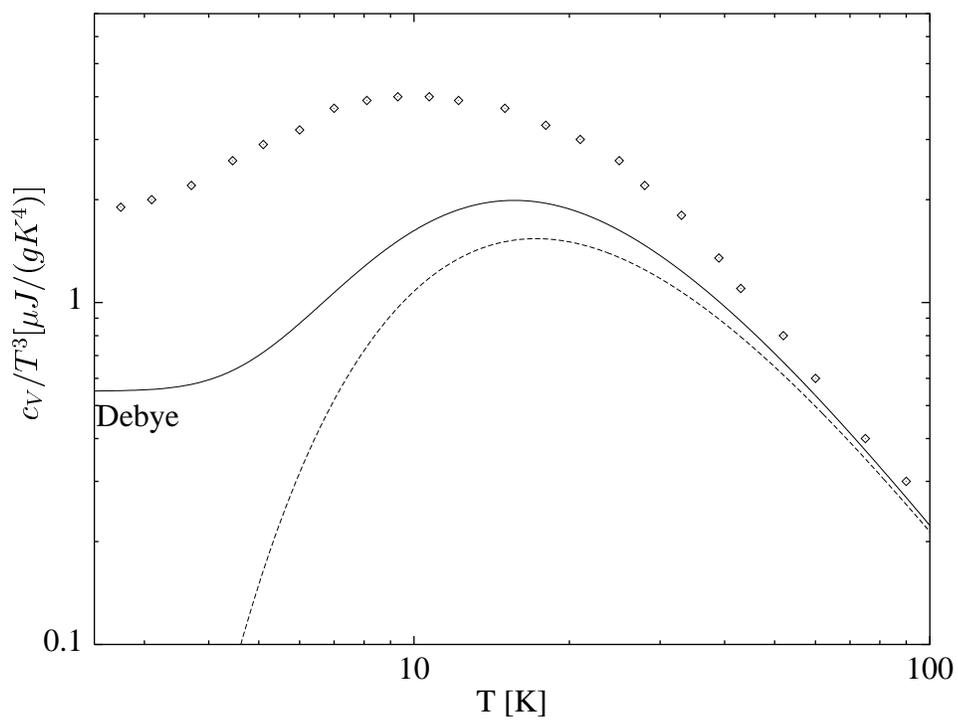
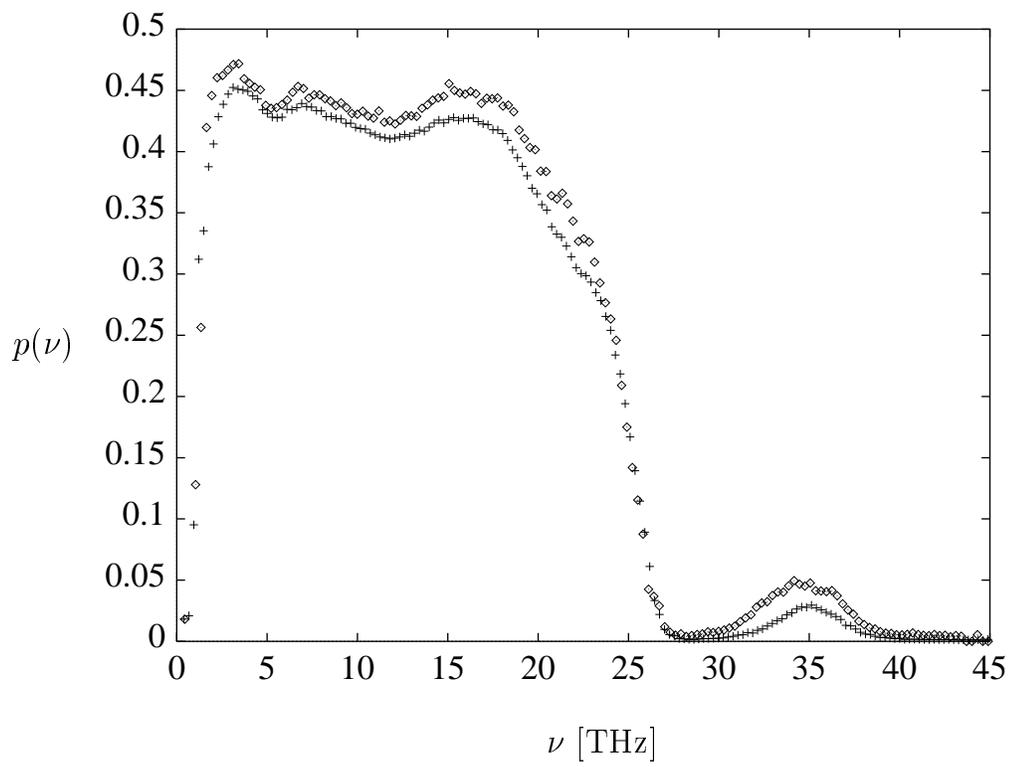
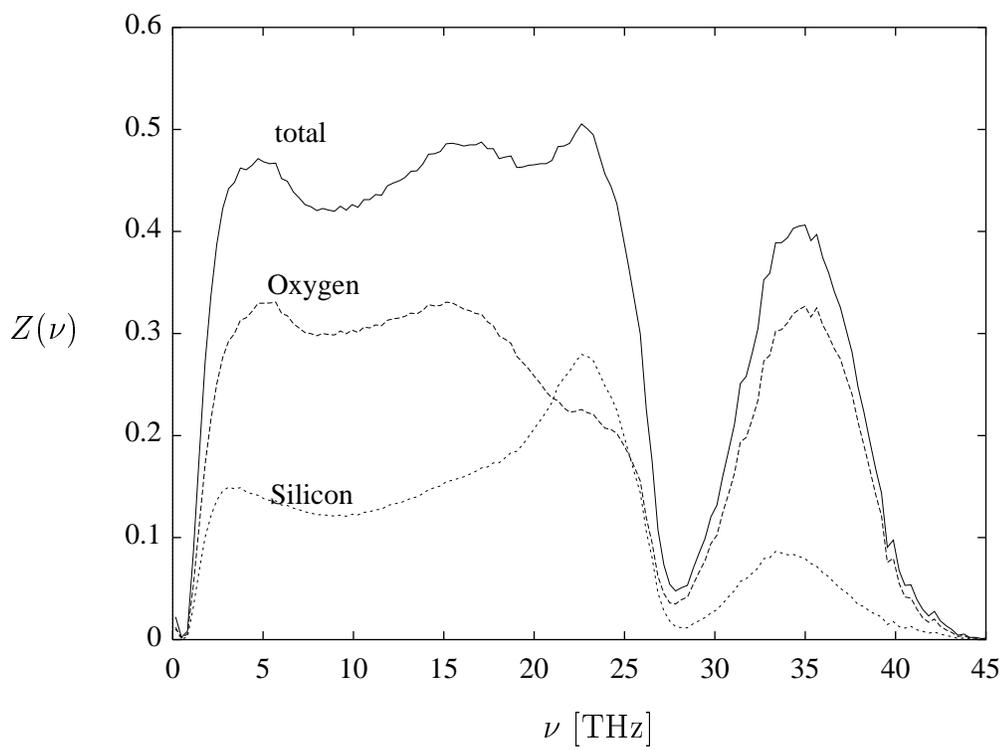
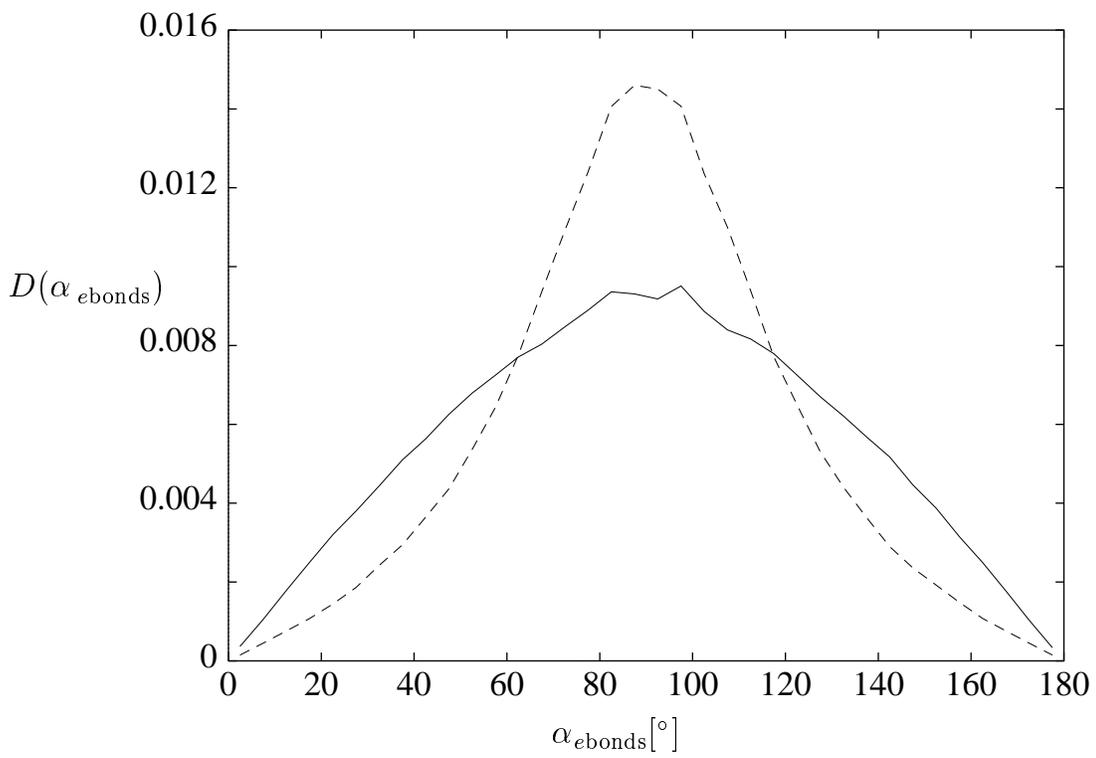
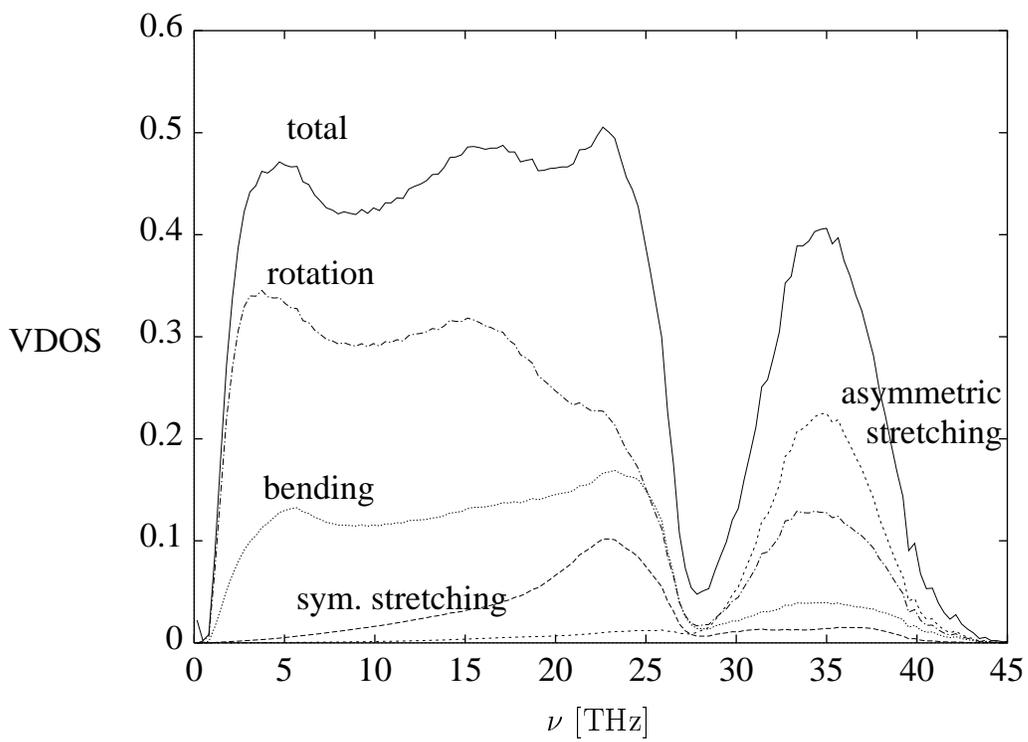
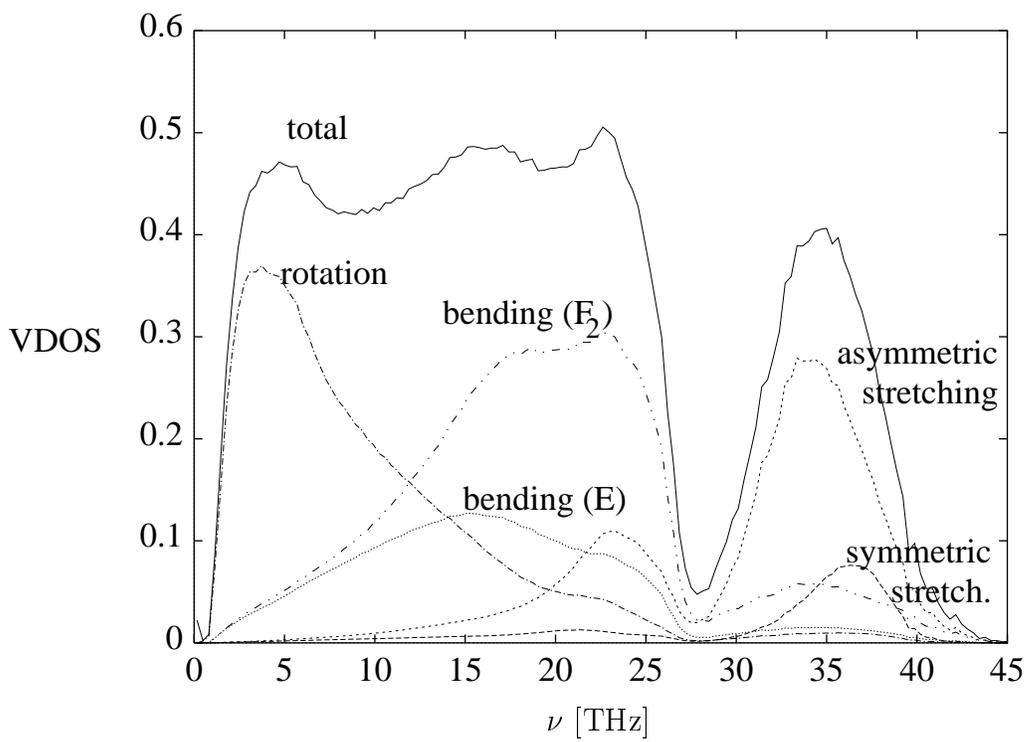
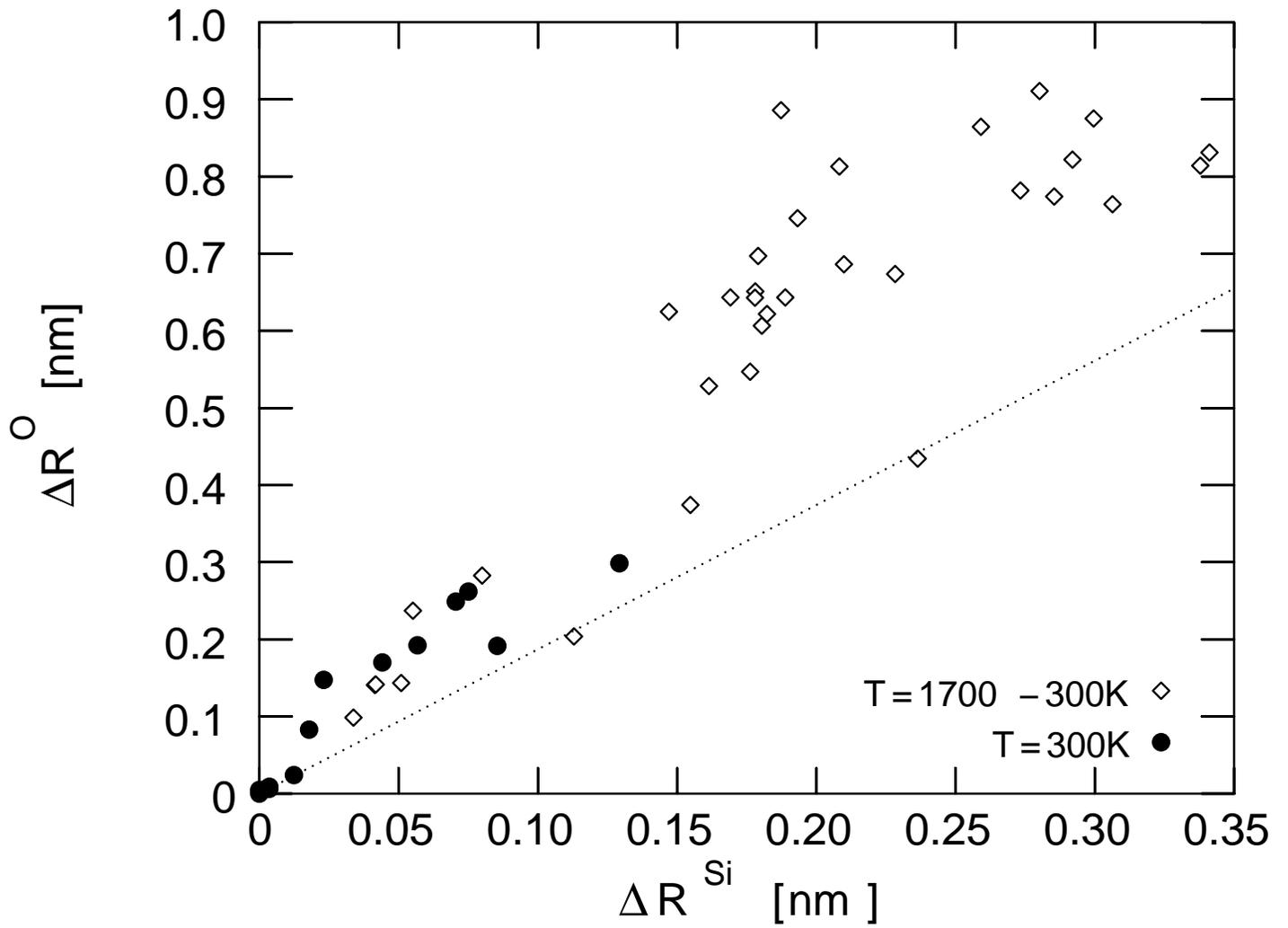
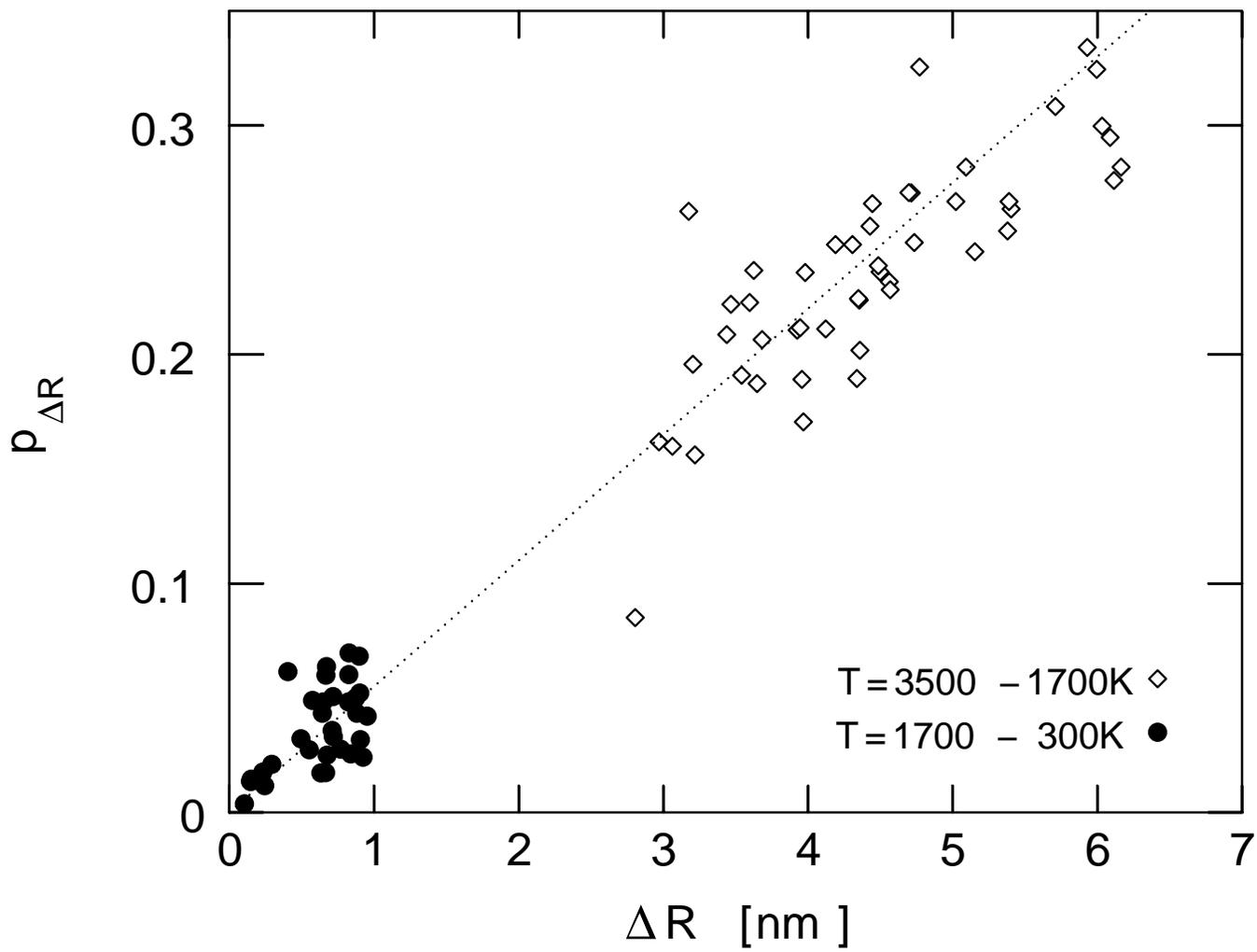
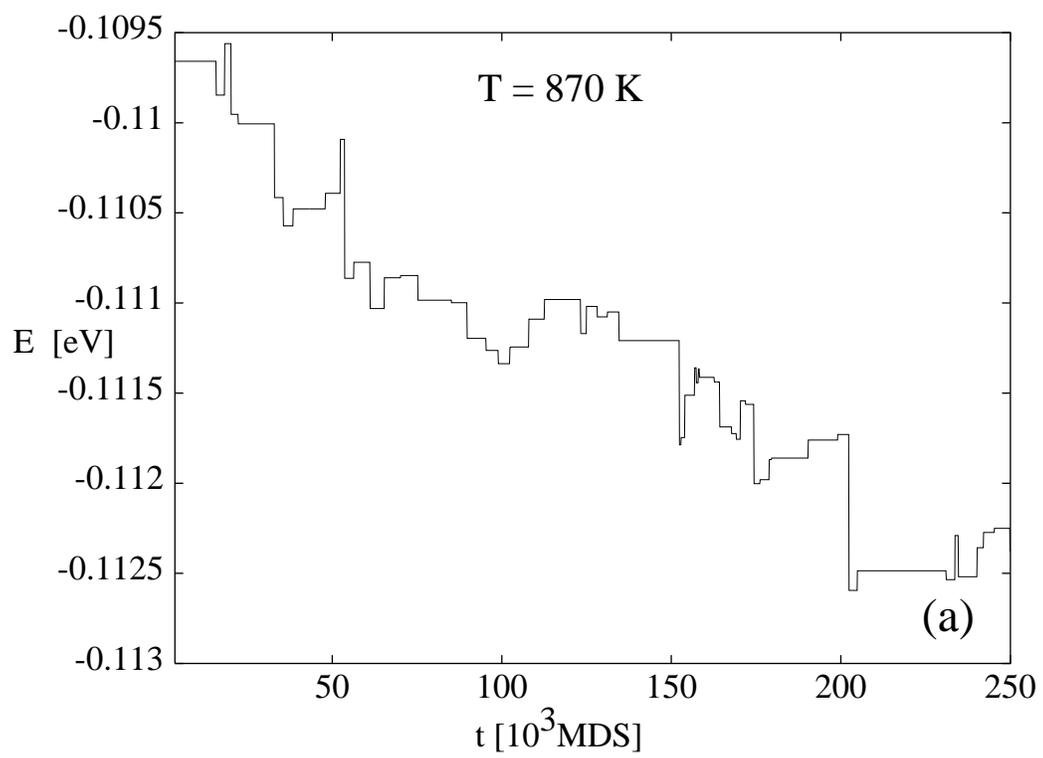
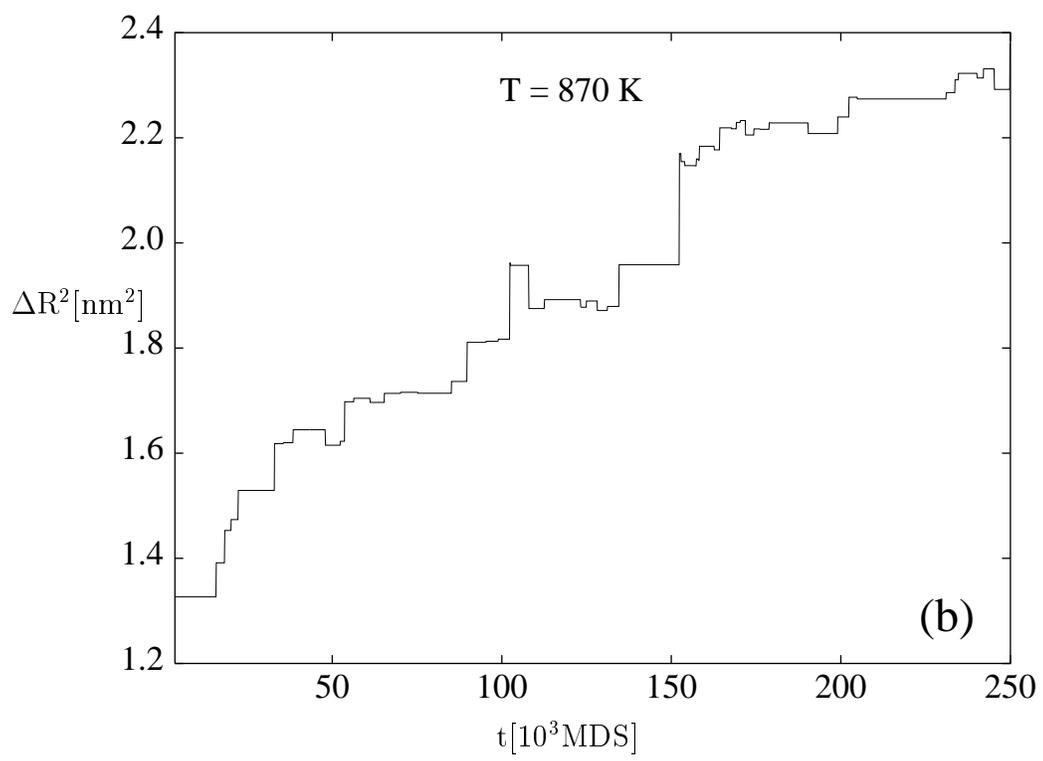
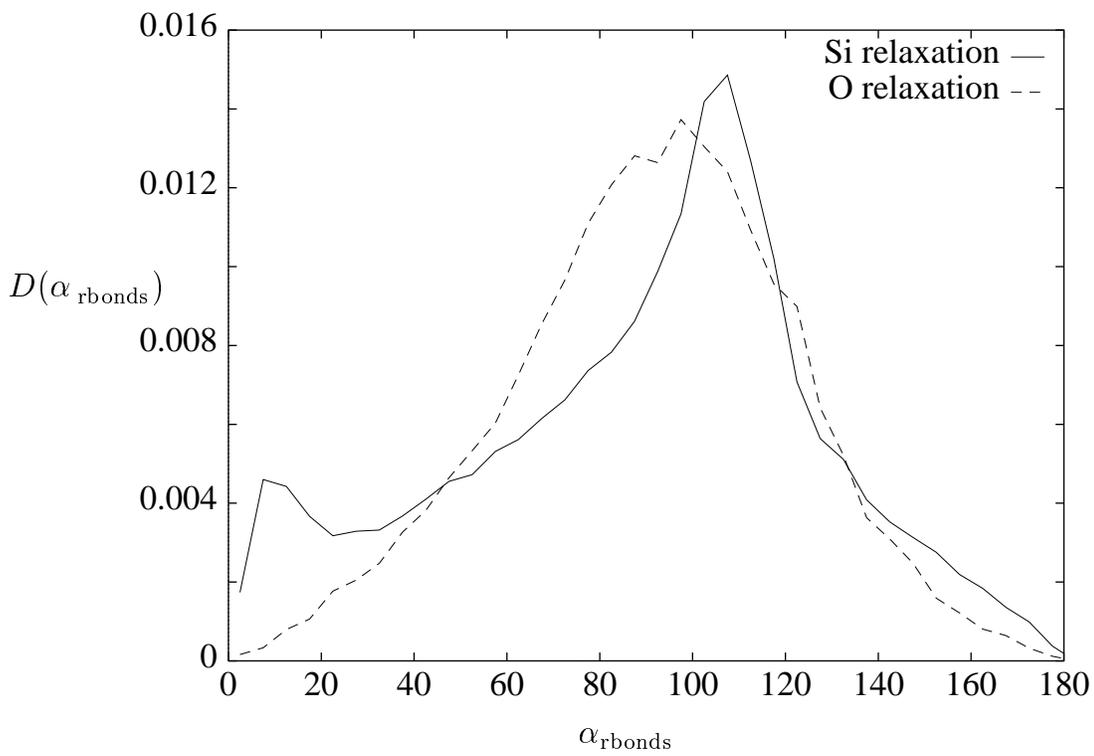
\end{document}